\documentclass[useAMS,usenatbib]{mn2e}
\usepackage{graphicx}
\usepackage{amssymb}
\usepackage{lscape}
\usepackage{ulem}
\usepackage{txfonts}

\def\kms{km~s$^{-1}$}
\def\cm{cm$^{-2}$}
\def\lya{Ly$\alpha$}
\def\nhi{$N$(H\,{\sc i})}
\def\hi{H\,{\sc i}}
\def\si2{Si\,{\sc ii}}
\def\c4{C\,{\sc iv}}
\def\mg2{Mg\,{\sc ii}}
\def\n5{N\,{\sc v}}
\def\fe2{Fe\,{\sc ii}}
\def\al2{Al\,{\sc ii}}
\def\zn2{Zn\,{\sc ii}}
\def\c2s{C\,{\sc ii}$^{\star}$}
\def\hkpc{$h_{70}^{-1}$ kpc}

\title[The nature of proximate DLAs]{The nature of proximate damped
Lyman alpha systems.\thanks{Based on observations made with ESO
Telescopes at the Paranal Observatories under program 080.A-0014(A).}}

\author[Ellison et al.] {Sara L. Ellison$^1$,  J. Xavier Prochaska$^2$,
 Joseph Hennawi$^3$, Sebastian Lopez$^4$, 
\newauthor Christopher Usher$^1$, Arthur M. Wolfe$^5$, David M. Russell$^6$, 
Chris R. Benn$^{7}$\\
$^1$ Department of Physics and Astronomy, University of Victoria, 
Victoria, British Columbia, V8P 1A1, Canada.\\
$^2$ Department of Astronomy and Astrophysics, UCO/Lick Observatory, University of California, 1156 High Street, Santa Cruz, CA 95064, USA \\
$^3$ Department of Astronomy, 601 Campbell Hall, University of California, Berkeley, CA 94720-3411, USA \\ 
$^4$ Departamento de Astronom\'ia, Universidad de Chile, Casilla 36-D, Santiago, Chile\\
$^5$ Department of Physics, and Center for Astrophysics and Space Sciences, University of California, San Diego, 9500 Gilman Drive, La Jolla, CA 92093-0424, USA\\
$^6$ Astronomical Institute `Anton Pannekoek', University of Amsterdam, P.O. Box 94249, 1090 GE Amsterdam, Netherlands\\
$^7$ Isaac Newton Group, Apartado 321, E-38700 Santa Cruz de La Palma, Spain
}

\begin{document}

\maketitle

\begin{abstract}

We present high resolution echelle spectra of 7 proximate damped Lyman
alpha (PDLA) systems whose relative velocity separation from the
background quasar is $\Delta V < 3000 $ \kms.  Combining our sample
with a further 9 PDLAs from the literature we compare the chemical
properties of the proximate systems with a control sample of
intervening DLAs.  The PDLAs are usually excluded from statistical
studies of absorption selected galaxies and this sample constitutes
the first systematic study of their chemical and ionization
properties.  Taken at face value, the sample of 16 PDLAs exhibits a
wide range of metallicities, ranging from $Z \sim 1/3 Z_{\odot}$ down
to $Z \sim 1/1000 Z_{\odot}$, including the DLA with the lowest
N(SiII)/N(HI) yet reported in the literature.  However, some of these
abundances may require ionization corrections. We find several pieces
of evidence that indicate enhanced ionization and the presence of a
hard ionizing spectrum in PDLAs which lead to properties that contrast
with the intervening DLAs, particularly when the \nhi\ is low.  The
abundances of Zn, Si and S in PDLAs with log \nhi\ $>$ 21, where
ionization corrections are minimized, are systematically higher than
the intervening population by a factor of around 3.  We also find
possible evidence for a higher fraction of NV absorbers amongst the
PDLAs, although the statistics are still modest.  6/7 of our echelle
sample show high ionization species (SiIV, CIV, OVI or NV) offset by
$>$100 \kms\ from the main low ion absorption.  We analyse
fine-structure transitions of CII$^{\star}$ and SiII$^{\star}$ to
constrain the PDLA distance from the QSO.  Lower limits range from
tens of kpc up to $>$160 kpc for the most stringent limit.  We
conclude that (at least some) PDLAs do exhibit different
characteristics relative to the intervening population out to 3000
\kms\ (and possibly beyond).  Nonetheless, the PDLAs appear distinct
from lower column density associated systems and the inferred
QSO-absorber separations mean they are unlikely to be associated with
the QSO host.  No trends with $\Delta V$ are found, although this
requires a larger sample with better emission redshifts to confirm.
We speculate that the PDLAs preferentially sample more massive
galaxies in more highly clustered regions of the high redshift
universe.

\end{abstract}


\section{Introduction}

Absorption systems with $13 <$ log \nhi\ $<$ 17 and within $\sim$ 5000
\kms\ of a background QSO are often considered to be `associated' with
the quasar host galaxy, its immediate environment or its ejecta.
Evidence for this conclusion comes from time variability of the
absorption (Hamann et al. 1997a; Wise et al. 2004; Lundgren et
al. 2007), partial coverage of the emission source (Hamann et
al. 1997b,c; Fox et al. 2008), elevated incidence of highly
ionized species (e.g. Ganguly et al. 2001; Richards 2001; Fechner \&
Richter 2009), possible dependence of absorber incidence on QSO
properties such as radio loudness or orientation (M\o ller \& Jakobsen
1987; Richards et
al. 2001; Baker et al. 2002; Vestergaard 2003; Wild et al. 2008) and
solar or super-solar metallicities (e.g. Petitjean, Rauch \& Carswell
1994; Hamann et al. 1997a; Srianand \& Petitjean 2000; D'Odorico et
al. 2004; Fechner \& Richter 2009).  The associated population of 
absorbers appears to extend out
to at least 10,000 \kms\ (Petitjean et al. 1994; Richards et al. 1999;
Wild et al. 2008; Tytler et al. 2009).

Although the $z_{\rm abs} \sim z_{\rm em}$ \lya\ forest, CIV, MgII and
Lyman limit systems have been well-researched in the last 20 years,
the same is not true of the proximate damped Lyman alpha systems (DLAs).  This
is partly a historical bias.  DLAs are excellent tools for
characterising the high redshift galaxy population.  Knowing that
associated systems may be connected with the QSO or its host galaxy,
statistical samples of DLAs have therefore excluded these potentially
special systems (e.g. Wolfe et al. 1995; Ellison et al. 2001; Prochaska,
Herbert-Fort \& Wolfe 2005; Jorgenson et
al. 2006; Noterdaeme et al. 2009).  Furthermore, the chemical
abundances of intervening DLAs can be determined fairly easily because 
their gas
is predominantly neutral.  Proximity to a QSO could invalidate that
assumption and lead to erroneous abundance determinations, unless
photoionization modelling is used.  However, it is for exactly these
reasons that the study of proximate DLAs (PDLAs) is interesting.  
PDLAs provide a potential probe of galaxies clustered
around (or associated with) QSOs, and even provide insight into the
interplay between ionizing radiation and the ISM.

One of the difficulties in studying the proximate DLAs is the small
redshift path for each line of sight, which means that PDLAs are very rare.
For example, assuming the same number density as intervening systems
at $z = 3$, $\sim$ 100 QSOs would need to be surveyed to find just one
PDLA.  A second challenge is constructing a sample of PDLAs selected
with a homogeneous QSO-absorber velocity offset ($\Delta V$)
criterion.  Since the derived emission redshift of a QSO depends
critically on the emission lines used (e.g. Gaskell 1982; Tytler \&
Fan 1992), the value can be uncertain by over 1000 \kms\ ($\sim$ 3 Mpc
proper at $z=3$).  The first systematic study of PDLAs was conducted
with the Complete Optical and Radio Absorption Line System (CORALS)
Survey data by Ellison et al. (2002) who found a factor of 4 excess in
the number density of PDLAs in their radio-selected sample.  Russell,
Ellison \& Benn (2006) extended this work to identify 33 PDLAs in the
Sloan Digital Sky Survey (SDSS) Data Release 3 (DR3) and confirmed the
PDLA excess, independent of radio loudness, at 3.5$\sigma$
significance.  In the largest PDLA survey conducted so far, Prochaska,
Hennawi \& Herbert-Fort (2008) use the SDSS DR5 to identify 108 PDLAs
with $\Delta V <$ 3000 \kms.  In addition to its larger size, this
work re-calculated the QSO redshifts in order to obtain improved
estimates of $\Delta V$.  Prochaska et al. (2008b) found that PDLAs
outnumber the intervening DLAs at $z \sim 3$ by a factor of 2, but
found no statistically significant excess at $z<2.5$ or $z>3.5$.

In addition to the PDLA number density, Prochaska et al. (2008b)
investigated whether their \hi\ column density distribution function
differed from the intervening DLAs.  They found marginal evidence for
an excess of high \nhi\ PDLAs.  Previously, Tytler (1982) had found no
excess for the slightly lower \nhi\ Lyman limit systems but with a much
larger sample Prochaska,
O'Meara \& Worseck (2010) find a deficit.  Apart from
their incidence and \nhi\ distribution, few studies of PDLAs exist to
assess whether or not they are consistent with the intervening
population.  In particular, PDLAs are rarely targetted for echelle
spectroscopy, so little is known about their chemical enrichment.
Nonetheless, a small number of PDLAs have appeared in chemical
abundance studies (see Table \ref{lit_z_table} for those that meet our
criteria) and they do not appear to have strikingly different
properties to the intervening DLAs (Lu et al. 1996; Akerman et
al. 2005; Rix et al. 2007).  The range of metallicities, $\alpha$
element enhancements and dust depletions are all consistent with the
range seen amongst intervening DLAs.  Rix et al. (2007) also found
negligible ionization corrections were necessary for the PDLA in their
study, as is usually the case for other DLAs.  It has therefore
usually been concluded that the PDLAs do not constitute a distinct
population from the intervening systems (e.g. M\o ller et al. 1998).

In this work, we present echelle spectra for seven PDLAs (Section
\ref{obs_sec}), determine improved emission redshifts (Section
\ref{z_sec}) and measure column densities for a range of metal line
species (Section \ref{coldens_sec}).  We supplement our sample with
other PDLAs critically selected from the literature to form a sample
of 16 proximate absorbers (Section \ref{lit_sec}).  Ionization
properties and metal abundances of PDLAs are compared to the
intervening systems in Sections \ref{ion_sec} and \ref{abund_sec}.
Finally, we use measurements of CII$^{\star}$ and upper limits to
SiII$^{\star}$ to constrain the distance between the PDLA and the
background QSO (Section \ref{dist_sec}).

Unless otherwise stated, we assume $\Omega_M = 0.3$, $\Omega_{\Lambda}=0.7$
and H$_0$ = 70 \kms\ Mpc$^{-1}$.

\section{Sample, observations and data reduction}\label{obs_sec}

Proximate DLAs for this study were mostly selected from the SDSS DR5
catalogue of Prochaska, Hennawi \& Herbert-Fort (2008).  These are
J014049.18$-$083942.5, J014214.74+002324.3, J113130+604420,
J124020.9+145535.6, J160413.9+395121.9 and J232115.48+142131.5.  For
simplicity we hereafter refer to these QSOs by their SDSS names in the
form Jhhmm+ddmm.  One additional non-SDSS QSO is also in the sample:
Q0151+048 (UM144).  SDSS magnitudes are given in Table \ref{obs_table}
for all QSOs except Q0151+048 where we have taken the R-band magnitude
from the APM\footnote{www.ast.cam.ac.uk/$\sim$apmcat} plate
measurements.  Our target selection was based on relatively bright
QSOs with absorbers whose velocity offset was found to be $\Delta V <
3000$ \kms\ by Prochaska et al. (2008b), and with positions accessible
during our scheduled observing runs.  The $\Delta V$ cut-off is
somewhat arbitrary and it should be stressed that it is unlikely that
these velocities correspond to distances inferred from the Hubble
expansion (d=v/H(z)) if the proximate absorbers are linked either to
the host galaxy or associated with its gravitational potential.  In
the former case, the velocities may reflect outflows or internal
motions.  In the latter scenario, the $\Delta V$ distribution would be
indicative of peculiar motions.  We provide empirical evidence that
the value of $\Delta V$ does not correspond to a Hubble distance by
showing that a number of the PDLAs have negative velocities relative
to the QSO (Section \ref{z_sec}) and that elemental trends do not
correlate with $\Delta V$ (Section \ref{ion_sec}).

Observations were obtained with either the UV-Visual Echelle
Spectrograph (UVES) at the Very Large Telescope (VLT) or with the High
Resolution Echelle Spectrograph (HIRES) on the Keck telescope.   

\subsection{UVES data acquisition and reduction}

UVES is a dual-arm echelle spectrograph with a grating cross
disperser.  There are four different cross dispersers available and
two dichroics (DIC1 and DIC2), which split the incident beam between the
blue and red arms.  The central wavelengths are governed by the grating
angles and are adjusted to allow the desired spectral coverage to fall
on the 3 CCDs (one in the blue arm and two in the red).  We combined
two of UVES's `standard' settings (DIC1 346+580 and DIC2 437+860) in
order to obtain spectral coverage over approximately 3050 -- 9700 \AA.
Integration times are given in Table \ref{obs_table}.  Observations of
a ThAr lamp for wavelength calibration followed each exposure.  The
CCDs were binned 2x2 and a 1 arcsecond slit was used, resulting in 
nominal resolutions of R=41,000 and 39,000 in the blue and red arms,
respectively.  We observed J0140$-$0839, J0142+0023 and J2321+1421 on
the nights of 2007 October 2-3 with UVES in visitor mode. The
seeing ranged between 0.6 and 1.0 arcsec.

The spectra were extracted and reduced using the UVES pipeline. The
extracted spectra from each exposure were converted to a common
vacuum-heliocentric wavelength scale.  Where multiple exposures in a
given setting were obtained, or where there is spectral overlap
between settings, the data were combined, weighting by the inverse of
the flux variances.  Table \ref{obs_table} lists the typical S/N
ratios for each target in the final combined spectrum (where there is
some wavelength overlap between settings).  The spectrum was
normalized by fitting a smooth continuum through unabsorbed parts of
the data using the Starlink DIPSO
package\footnote{http://star-www.rl.ac.uk/star/docs/sun50.htx/sun50.html}.

\subsection{HIRES data acquisition and reduction}

J1240+1455 and J1604+3951 were observed with the Keck/HIRES
spectrometer (Vogt 1994) using the upgraded CCD mosaic on 2007 April
27--28.  J1131+6044 and Q0151+048 were also observed with HIRES on
2006 December 25 and 2006 August 17 respectively.  The first two
quasars, which have PDLAs at $z_{\rm abs} < 3$, were observed using
the blue cross-disperser (HIRESb) with wavelength coverage from
roughly the atmospheric cutoff to $\approx 6000$\AA.  The latter two
were observed using the red cross-disperser (HIRESr) with the kv418
blocking filter giving an effective wavelength coverage of
4300--8700\AA.  The echelle and cross disperser angles are given in
Table \ref{obs_table}, along with total exposure times.  All of the
observations were acquired through the C1 decker which affords a full
width at half maximum (FWHM) ~$\approx 6 \, \rm km \, s^{-1}$ spectral
resolution.  The data were reduced with the HIRedux pipeline that is
available within the XIDL software
package\footnote{http://www.ucolick.org/$\sim$xavier/IDL}.  One
significant difference in the analysis of the HIRES data, compared to
the UVES reductions, is the stage at which the the continuum is
fitted.  Whereas the UVES data are normalized as the final reduction
step once the echelle orders have been combined into one dimension,
the XIDL reduction pipeline fits a local continuum to each order
before the 2-dimensional spectra are combined into 1D.  The advantage
of this latter approach is that the construction of the 1-dimensional
spectrum is simplified, e.g. by removing the effects of the blaze.
The disadvantage is that since damped \lya\ troughs
extend over multiple orders, HIRES spectra do not lend themselves
easily to \lya\ fitting.  We discuss this further below.

\begin{center}
\begin{table*}
\begin{tabular}{lcccccccc}
\hline
QSO & $r$ mag &  $z_{\rm em}$  &  $z_{\rm abs}$  & log \nhi\ (\cm) & Instrument & Instrument  & Exposure & S/N  \\
    &  (SDSS) &                &                 & (UVES/SDSS)   & &  set-up & time (s) & pix$^{-1}$ \\ \hline 
J0140$-$0839 & 17.7 & 3.7156   & 3.6960 & 20.75$\pm$0.15 & UVES &  346+580  & 6000 & 30 -- 50\\
             &      &          &        &                &       & 437+860  & 6000 & \\
J0142$+$0023 & 18.3 & 3.3734   & 3.34765& 20.38$\pm$0.05 & UVES &  346+580  & 6000 & 30 -- 50 \\
             &      &          &        &                &  &      437+860  & 10200 & \\
Q0151+048    & 17.5 & 1.9225   & 1.9342 & 20.34$\pm$0.02 &  HIRES & 0/1.250   & 9000  & 15 -- 20\\
J1131+6044   & 17.7 & 2.9069   & 2.8754 & 20.50$\pm$0.15 &  HIRES & 0/1.680  & 7200  & 15 -- 20  \\
J1240+1455   & 18.9 & 3.1092   & 3.1078 & 21.3$\pm$0.2   &  HIRES & $-0.110$/0.557 & 5400   & 5 -- 8\\
J1604+3951   & 18.1 & 3.1542   & 3.1633 & 21.75$\pm$0.2  &  HIRES & $-0.110$/0.557  & 10300  & 10 -- 15\\
J2321+1421   & 18.3 & 2.5539   & 2.5731 & 20.70$\pm$0.05 &   UVES &346+580  & 3000 & 10 -- 30\\
             &      &          &        &                &  &     437+860  & 11700 & \\
\hline 
\end{tabular}
\caption{\label{obs_table}Target list and observing journal.  The
instrument set-up for HIRES refers to the echelle/cross-disperser
angles.  The S/N is given as a representative value of the final
combined spectra of both settings.}
\end{table*}
\end{center}

\section{Redshifts}\label{z_sec}

\begin{center}
\begin{table*}
\begin{tabular}{lcccccc}
\hline
QSO           & log N(HI) & $z_{\rm em}$  &  $z_{\rm abs HI}$ &   $z_{\rm abs Z}$  & $\Delta$V$_{\rm HI}$ &  $\Delta$V$_{\rm Z}$ \\ 
 & (\cm) & & & & (\kms) & (\kms)  \\
\hline 
J0140$-$0839  & 20.75$\pm$0.15 & 3.7156$\pm$0.012 & 3.6960 & 3.69660 & 1250 & 1211  \\
J0142$+$0023  & 20.38$\pm$0.05 & 3.3734$\pm$0.008 & 3.34765& 3.34768 & 1772 & 1769 \\
Q0151+048 & 20.34$\pm$0.02 & 1.9225$\pm$0.003 & 1.9342 & 1.93429 & $-1199$ & $-1208$ \\
J1131+6044 & 20.50$\pm$0.15 & 2.9069$\pm$0.009 & 2.8754 & 2.87562 & 2424 & 2412 \\
J1240+1455 & 21.3$\pm$0.2 & 3.1092$\pm$0.005   & 3.1078 & 3.10803 & 102 & 85  \\
J1604+3951 & 21.75$\pm$0.2 & 3.1542$\pm$0.007   & 3.1633 & 3.16711 & $-656$ & $-930$  \\
J2321+1421 & 20.70$\pm$0.05 & 2.5539$\pm$0.006 & 2.5731 & 2.57312 & $-1616$ & $-1618$  \\
\hline 
\end{tabular}
\caption{\label{dv_table}PDLA redshifts and relative velocities}
\end{table*}
\end{center}

Accurate emission and absorption redshifts are required in order to
define a sample of PDLAs lying within a given velocity range $\Delta
V$ of the QSO.  Typical limiting values of $\Delta V$ adopted in the
literature, when defining samples of associated absorbers, are
$<$3000 to $<$6000 km/s.  $\Delta V$ is calculated as:

\begin{equation}
\Delta V = c\frac{R^2 - 1}{R^2 + 1}
\end{equation}

\noindent  where c is the speed of light and 

\begin{equation}
R \equiv \frac{1+z_{\rm abs}}{1+z_{\rm em}}
\end{equation}

\noindent where $z_{\rm abs}$ and $z_{\rm em}$ are the absorption and
emission redshifts respectively.  It is well known that the
determination of QSO emission redshift is sensitive to the choice of
emission lines (e.g. Gaskell 1982; Tytler \& Fan 1992), where [OIII]
$\lambda \lambda$ 4959, 5007 is usually considered one of the most
reliable indicators of systemic velocity.  The SDSS composite spectrum
of vanden Berk et al. (2001) shows that narrow forbidden lines may
only be shifted by $\pm$ tens of \kms, but the commonly used \lya, CIV
and CIII] lines may be offset by many hundred \kms.  The majority of
the PDLAs used in this study (see Section \ref{obs_sec}) are selected
from the SDSS DR5 survey of Prochaska et al. (2008b), and we adopt
their re-derived emission redshifts. For the QSOs not in the sample of
Prochaska et al. (2008b) we derive emission redshifts in an identical
way.  The technique first calculates line centres of all significantly
detected emission lines as described by Hennawi et al. (2006) and
applies the known systematic shifts of each species as tabulated by
Shen et al. (2007) and Richards et al. (2002).  The high ionization
lines such as CIV and SiIV give the poorest estimate of $z_{\rm em}$,
with a redshift error of $\sim$ 700 \kms.  From the SDSS spectra, our
best estimates of systemic redshift are derived from the MgII emission
line which yields $z_{\rm em}$ accurate to $\sim$ 270 \kms, based on
the distribution of MgII-[OIII] offsets given in Richards et al. (2002).  
Emission redshift errors in Table \ref{dv_table} are based on the
distributions of line shifts given in Table 7 of Shen et al. (2007). For
future studies, it would be desirable to obtain infra-red (IR) spectra
of these targets to obtain redshifts directly from the Balmer and
forbidden [OIII] lines (e.g. Rix et al. 2007), the latter of which
yields accuracies of less than 50 \kms.  Recently Hennawi et
al. (2009) have obtained an IR spectrum for one of our sightlines:
J1240+1455 and for this object we adopt the emission redshift
determined from the [OIII] line ($z_{\rm em} = 3.1092$).  This value
is larger by 160 \kms\ than the value derived from the SDSS spectrum
($z_{\rm em} = 3.1070$) and is now larger than the absorption
redshift.

Having made our best estimate of the emission redshifts of the QSOs in
our sample, a value of the absorption redshift must also be obtained.
The velocity spread of metal species in DLAs can extend over several
hundred \kms\ (e.g. Prochaska et al. 2008a) and the redshift of the
best fitting \lya\ line can be offset from the metals.  In Table
\ref{dv_table} we give the absorption redshifts of our sample
determined from both the \lya\ line and the strongest metal component
($z_{\rm abs HI}$ and $z_{\rm abs Z}$ respectively).  It can be seen
that the resulting difference in $\Delta V$ is much smaller than the
typical error associated with the emission redshift determination.
Since the \lya\ redshift is the value most easily recovered from the
literature, we adopt that as our fiducial measure of $z_{\rm abs}$.

\section{Column density measurements}\label{coldens_sec}

\begin{figure}
\centerline{\rotatebox{0}{\resizebox{8cm}{!}
{\includegraphics{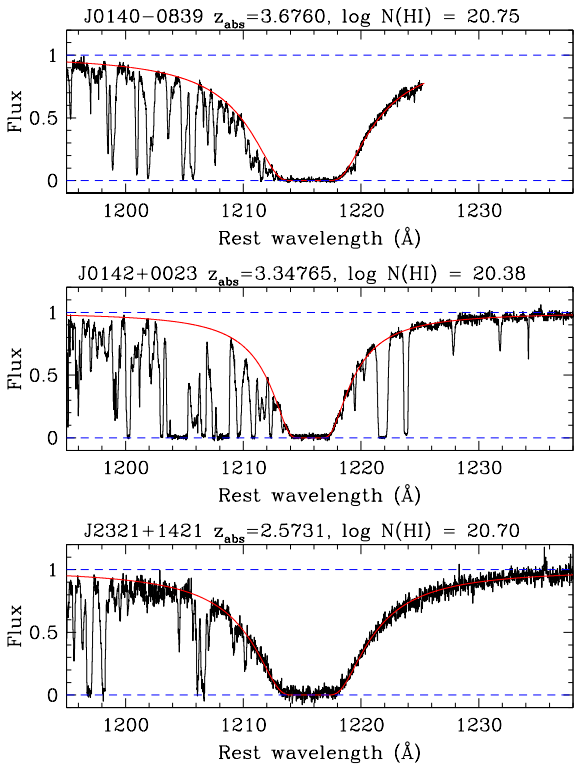}}}}
\caption{\label{HI_fits_PDLA} Fits to the \lya\ profiles for the three
PDLAs in our sample observed with UVES.  Column densities
and uncertainties are given in Table \ref{obs_table}. }
\end{figure}

For the 3 targets with UVES data, \hi\ column densities were derived
by fitting damped profiles to the \lya\ absorption in the normalized
spectra using the DIPSO software package.  Continuum adjustment and
\hi\ fitting were carried out iteratively, due to proximity of the
absorption line to the strong \lya\ emission of the QSO itself.  The
\hi\ column densities derived from the UVES data are all in good
agreement with those derived from the SDSS spectra published in
Prochaska et al. (2008b).  Figure \ref{HI_fits_PDLA} shows the \lya\
fits to the PDLAs.  For 2/3 of the UVES targets we were able to check
our \lya\ fits against an independent fit of Ly$\beta$ (for
J0142$+$0023 Ly$\beta$ is blended with strong \lya\ at a lower
redshift).  In both cases the \lya\ and Ly$\beta$ column densities
agree to within the errors quoted in Tables \ref{obs_table} and
\ref{dv_table}.

  For the 4 HIRES targets where DLA fitting is hampered by the
order-by-order normalization in XIDL, we adopt the HI column density
derived from the SDSS spectrum (Prochaska et al. 2008b).  As described
above, the fits to the UVES spectra demonstrate that the SDSS values
agree with the higher resolution spectra fits to within the errors.
For Q0151+048, the \hi\ column density was measured from an X-shooter
spectrum by Zafar et al. (in preparation), which agrees with the value
of M\o ller et al. (1998) to within 0.02 dex.

Metal column densities were mostly derived by fitting Voigt profiles
with the \textsc{VPFIT}
package\footnote{http://www.ast.cam.ac.uk/\~{}rfc/vpfit.html}.  The
fitting of Voigt profiles is particularly useful in distinguishing
contamination from blended absorption features and simultaneously
fitting multiple transitions of a given species.  In a few cases
(noted below), usually when the line was too weak, or the S/N too low
for a convincing Voigt profile decomposition, the apparent optical
depth method (AODM, e.g. Savage \& Sembach 1991) was applied.  In
general, the kinematic absorption model parameterized by redshifts and
$b$-values (Doppler widths) for multiple components was derived
independently for each species.  Again, exceptions to this are noted
below, usually in cases where the lines are weak or blending is
suspected.  Tables \ref{lowion_table} and \ref{highion_table} list the
adopted column density for low and high ionization species
respectively.

The calculation of upper limits depends on an assumption of the number
of pixels over which a line should be detected.  For a line of given
FWHM in \AA, the observed frame $n$ sigma
equivalent width (EW) detection limit is given by

\begin{equation}\label{EW_eqn}
EW = \frac{n \times FWHM}{S/N}
\end{equation}

\noindent where S/N is the signal-to-noise per pixel.  In order to
determine the appropriate value of the FWHM, we take the $b$-value of
the strongest component of a detected species for a given line of
sight (this is usually a transition of SiII or FeII) and apply the
correction FWHM=$b \times 2 \sqrt{ln2}$, before converting from \kms\
into \AA.  A single value of the FWHM is used for all low ions,
CII$^{\star}$ and AlIII limits in a given sightline.  Typical
$b$-values are 4--8 \kms.  If the strongest component is unresolved, a
FWHM corresponding to the instrumental resolution is adopted, i.e.
the FWHM that appears in Eqn \ref{EW_eqn} is the maximum of either
the strongest component or the instrumental resolution.  Had we
assumed that the lines were unresolved rather than adopting true line
widths, the upper limits would have been under-estimated by 0.2--0.3 dex.
For CIV, SiIV and NV we assume a broader profile and adopt a single
value of $b$= 10 \kms.  Upper limits in this paper are quoted at
3$\sigma$ significance.

For species where only saturated lines are detected, lower limit
column densities have been derived by applying the AODM to the
observed line with the lowest oscillator strength ($f$ value).  Column
densities are converted to abundances by adopting the solar scale of
Asplund et al. (2005), with the exception of argon for which we adopt
the improved value of Asplund et al. (2009).  Meteoritic values are
adopted with the exception of C, N and O where we take the
photospheric values (and Ar where several indirect methods are used,
see Asplund et al 2009 for details).  We note that there are
significant differences in the solar abundances of some of these
volatile elements relative to other commonly used reference scales,
e.g. Holweger (2001).  All literature abundances used for comparison
in the present work have therefore been re-calculated on the solar
scale of Asplund et al. (2005).  Following the standard procedure in
DLA abundances, we assume that the singly ionized species dominate the
total column densities (i.e. for element X, N(XII)=N(X)) and we do not
apply ionization corrections.  Even though this assumption may not
hold in some of the PDLAs in our sample (we discuss ionization in
Section \ref{ion_sec}), it is nonetheless of interest to apply this
common practice to the PDLAs to assess its impact.  Exceptions to this
approximation are three-fold for the
elements reported from our data.  Charge exchange reactions (and
ionization potentials $<$ 13.6 eV) keep the majority of nitrogen and
oxygen in the neutral state and for aluminium both AlII and AlIII can
contribute significantly.  Since we do not calculate ionization
corrections on a system-by-system basis, aluminium is omitted in our
final abundance listing, see Table \ref{abund_table}.

Abundances are quoted relative to solar using the usual notation

\begin{equation}\label{abund_eqn}
[X/H] = log \left[\frac{N(X)}{N(H)}\right] - log
\left[\frac{N(X)}{N(H)}\right]_{\odot}.
\end{equation}

\subsection{J0140$-$0839, log \nhi\ = 20.75}\label{0140_sec}

\begin{figure*}
\centerline{\rotatebox{270}{\resizebox{12cm}{!}
{\includegraphics{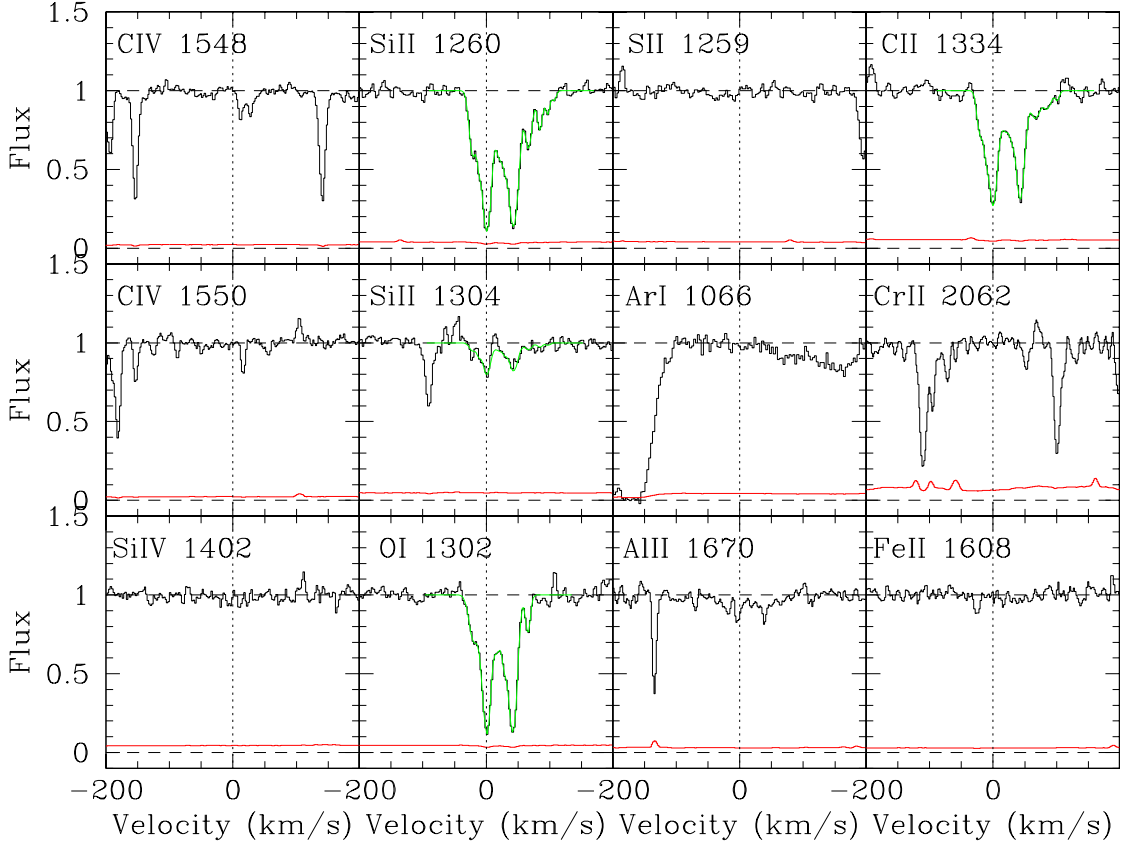}}}}
\caption{\label{0140_metals} Selected metal line transitions in the
PDLA towards J0140$-$0839.  The lower solid (red) line shows the error
array.  When fits have been derived using VPFIT, those fits are overlaid in
green.  $b$-values for this absorber range from 2 -- 8 \kms.  
Velocities are plotted relative to $z_{\rm abs}$=3.6966. }
\end{figure*}

The PDLA towards J0140$-$0839 has a simple velocity structure with two
main components in the low ionization species at $v=$0 and +50 \kms\
(see Figure \ref{0140_metals}).  Three unsaturated transitions of SiII
are detected ($\lambda_0=$1260, 1304, 1526 \AA), although the latter
of these exhibits some blending.  The column density is derived by
fitting the two bluer transitions simultaneously in \textsc{VPFIT}.
ArI $\lambda$ 1048 is blended but the redder transition (ArI $\lambda$
1066) of the doublet coincides with a clean region of the spectrum and
is used to determine an upper limit to the argon column density.
Despite the fairly high S/N (30 -- 50), AlII is barely detected and
its column density is derived using both \textsc{VPFIT} and the AODM.
In \textsc{VPFIT} we adopt the $b$-values and redshifts of the two
strongest components of the SiII fit to derive log
N(AlII)=11.78$\pm$0.04.  The AODM gives log N(AlII)=11.85$\pm$0.14, in
agreement with the fit, within the errors.  We adopt the average value
of these two methods.  No high ionization species were detected for
this PDLA. CIV is usually detected even in the low metallicity \lya\
forest (e.g. Cowie et al. 1995; Songaila 1997; Ellison et al. 1999,
2000), and at column densities 1--3 orders of magnitude above our
detection limit of N(CIV) $<$ 12.18.  However, this PDLA is extremely
metal-poor; both its Si and Fe abundances are the lowest in the
current published literature (of which we are aware).  Fox et al.
(2007) have shown that there is a broad correlation between N(CIV) and
metallicity, consistent with the non-detection of CIV in this PDLA.
As we discuss in Section \ref{ion_sec} there may be corrections due to
ionization effects, (and additionally some dust depletion, at least in
the case of Fe).  However, the OI is little affected by dust and
ionization and the OI $\lambda$ 1302 line is unsaturated in this PDLA,
from which we determine [O/H]=$-2.72$.  This supports an intrinsically
low metallicity for this system.  Indeed, the [O/H] of this PDLA is
similar to the oxygen abundance of the IGM (Simcoe et al. 2004;
Aguirre et al. 2008) who find [O/H] $\sim$ $-2.5$ to $-3$ depending on
the local over-density, although this depends on the model for
ionization corrections and shape of the UV background.  
Previously, a metallicity `floor' has been
suggested for DLAs (e.g. Prochaska et al. 2003a) and this system
offers a rare chance to measure abundances at extremely low levels of
enrichment.  For example, Akerman et al. (2005) and Pettini et
al. (2008) discuss how carbon abundances at low metallicity might
provide clues into the yields of early stellar populations.  The
unsaturated CII $\lambda$ 1334 yields [C/O] =$-0.29$ supporting the
idea of an upturn in C/O ratios at low (O/H) (Pettini et al. 2008).
The non-detection of N also sets a very low upper limit to N/O:
[N/O] $<-1.43$.

\subsection{J0142$+$0023, log \nhi\ = 20.38}\label{0142_sec}

\begin{figure*}
\centerline{\rotatebox{270}{\resizebox{12cm}{!}
{\includegraphics{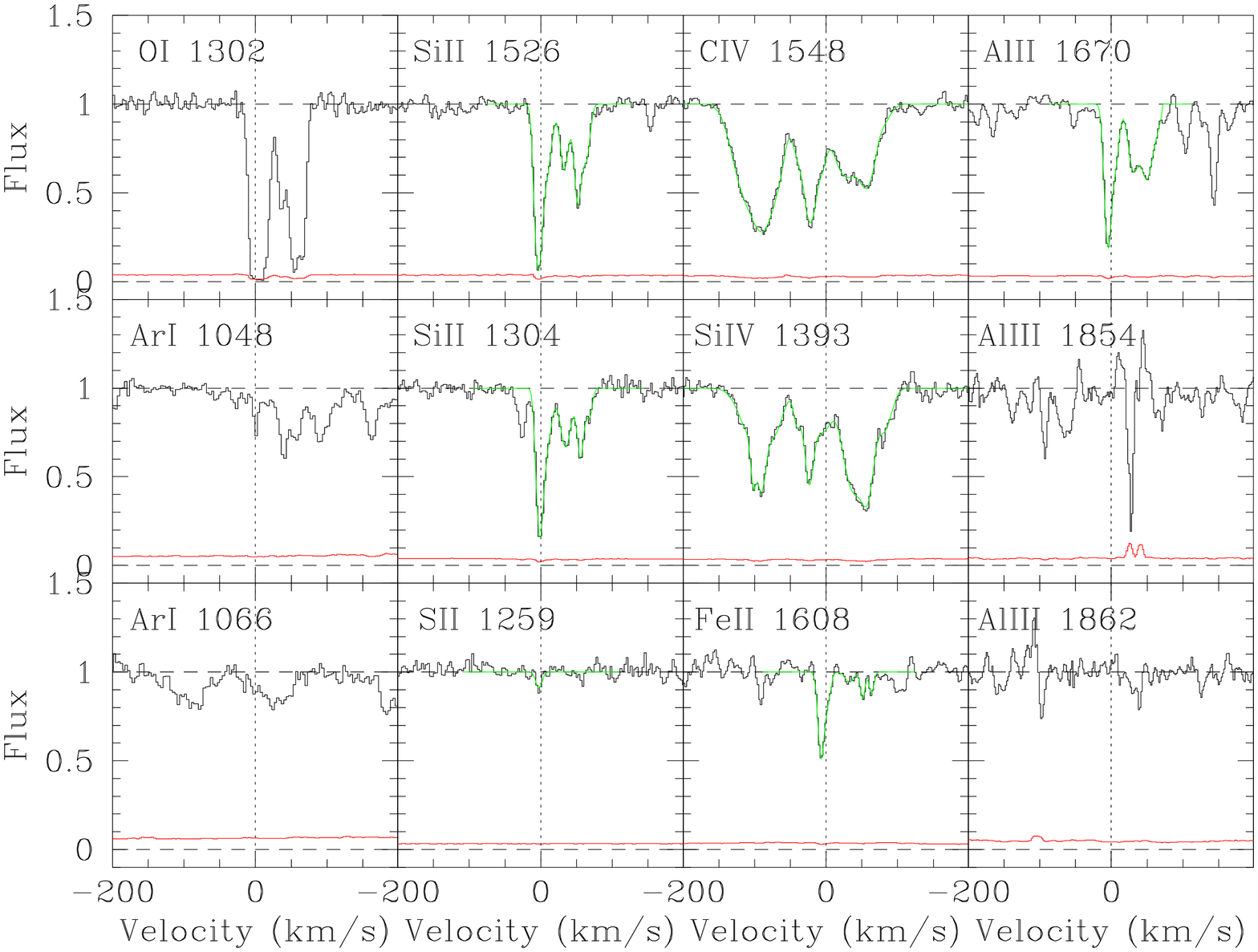}}}}
\caption{\label{0142_metals} Selected metal line transitions in the
PDLA towards J0142$+$0023.  The lower solid (red) line shows the error
array.  When fits have been derived using VPFIT, those fits are
overlaid in green.  $b$-values for this absorber range from 3 -- 9
\kms\ for the low ionization species and 7 -- 25 \kms\ for the higher
ionization species.  Velocities are plotted relative to $z_{\rm abs}$=
3.3477. }
\end{figure*}

Selected metal lines are shown in Figure \ref{0142_metals}.  Two lines
of SiII ($\lambda$ 1304 and $\lambda$ 1526) are detected, although
they are approaching saturation.  N(SiII) was obtained separately for
the two transitions with \textsc{VPFIT} and found to be in excellent
agreement.  AlII $\lambda$ 1670 follows the velocity structure of the
other low ions.  There is weak absorption at AlIII $\lambda$ 1862 at a
similar velocity, although gas in a different ionization state need
not necessarily trace the neutral gas.  A comparison with AlIII
$\lambda$ 1854 could confirm the detection, but unfortunately this
line is blended, so the detection of the $\lambda$ 1862 line remains
uncertain.  We therefore calculate N(AlIII) from AlIII $\lambda$1862
and take this as an upper limit.  SII $\lambda$ 1259, although weak,
is statistically a 5$\sigma$ detection.  Absorption is detected from
both SiIV and CIV.  Absorption is present in a similar velocity range
as the low ions, but with additional components present at negative
velocites from $-150$ to 0 \kms.  There is a significantly detected
feature at the expected velocity of ArI $\lambda$ 1048, but additional
absorption to the red alerts us to the possibility of contamination.
The identification of this line as ArI $\lambda$ 1048 could be
confirmed by the simultaneous detection of ArI $\lambda$ 1066, but
this line is blended with \lya.  Assuming the $v=0$ feature is ArI
$\lambda$ 1048, we derive a column density from the AODM and quote
this as a conservative upper limit.

\subsection{Q0151+048, log \nhi\ = 20.34}\label{0151_sec}

\begin{figure*}
\centerline{\rotatebox{270}{\resizebox{12cm}{!}
{\includegraphics{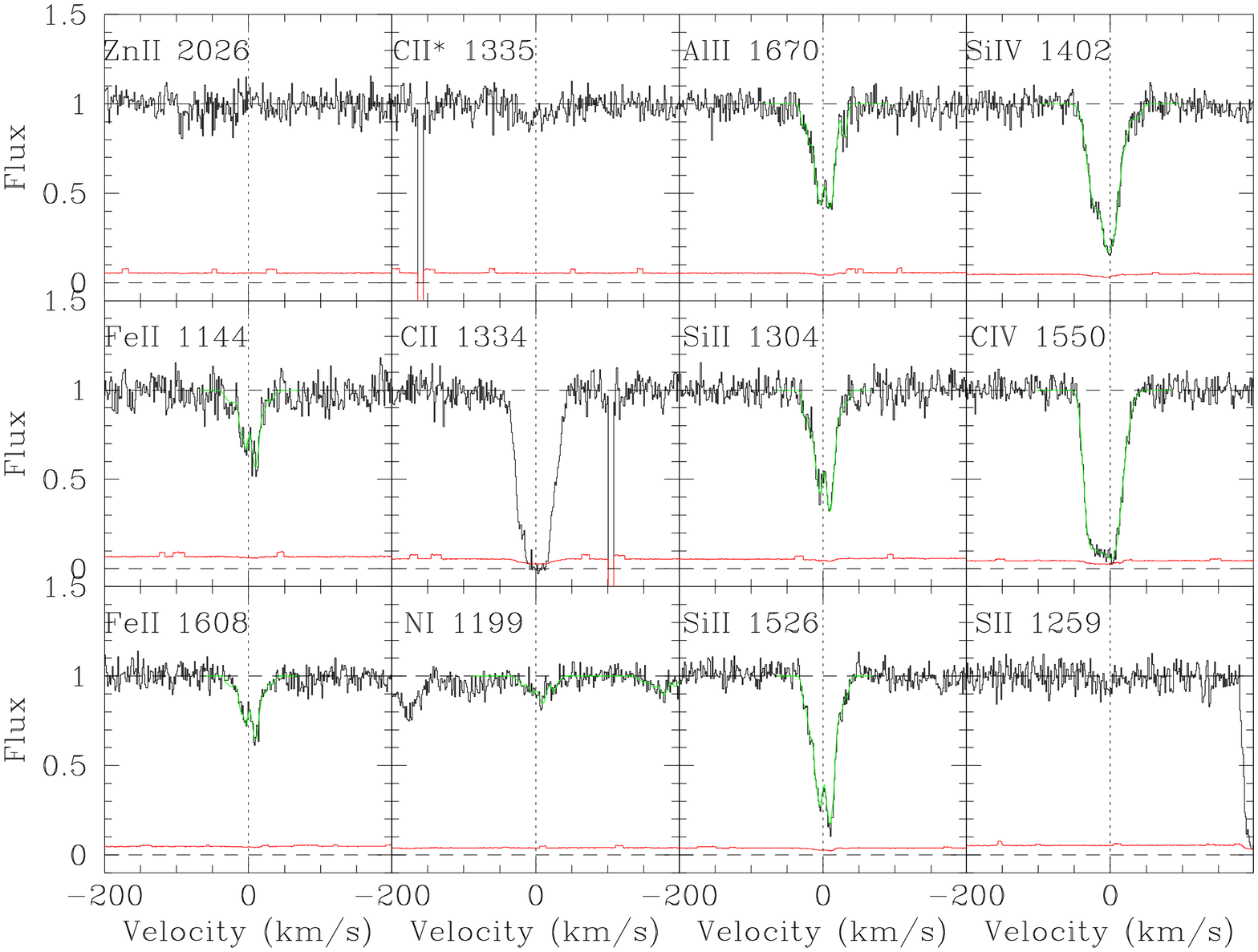}}}}
\caption{\label{0151_metals} Selected metal line transitions in the PDLA
towards Q0151+048.  The lower solid (red) line shows the error
array.  When fits have been derived using VPFIT, those fits are overlaid in
green.   $b$-values for this absorber range from 2 -- 8
\kms\ for the low ionization species and 5 -- 15 \kms\ for the higher
ionization species.  Velocities are plotted relative to $z_{\rm abs}$=
1.9342. }
\end{figure*}

\begin{figure}
\centerline{\rotatebox{0}{\resizebox{9cm}{!}
{\includegraphics{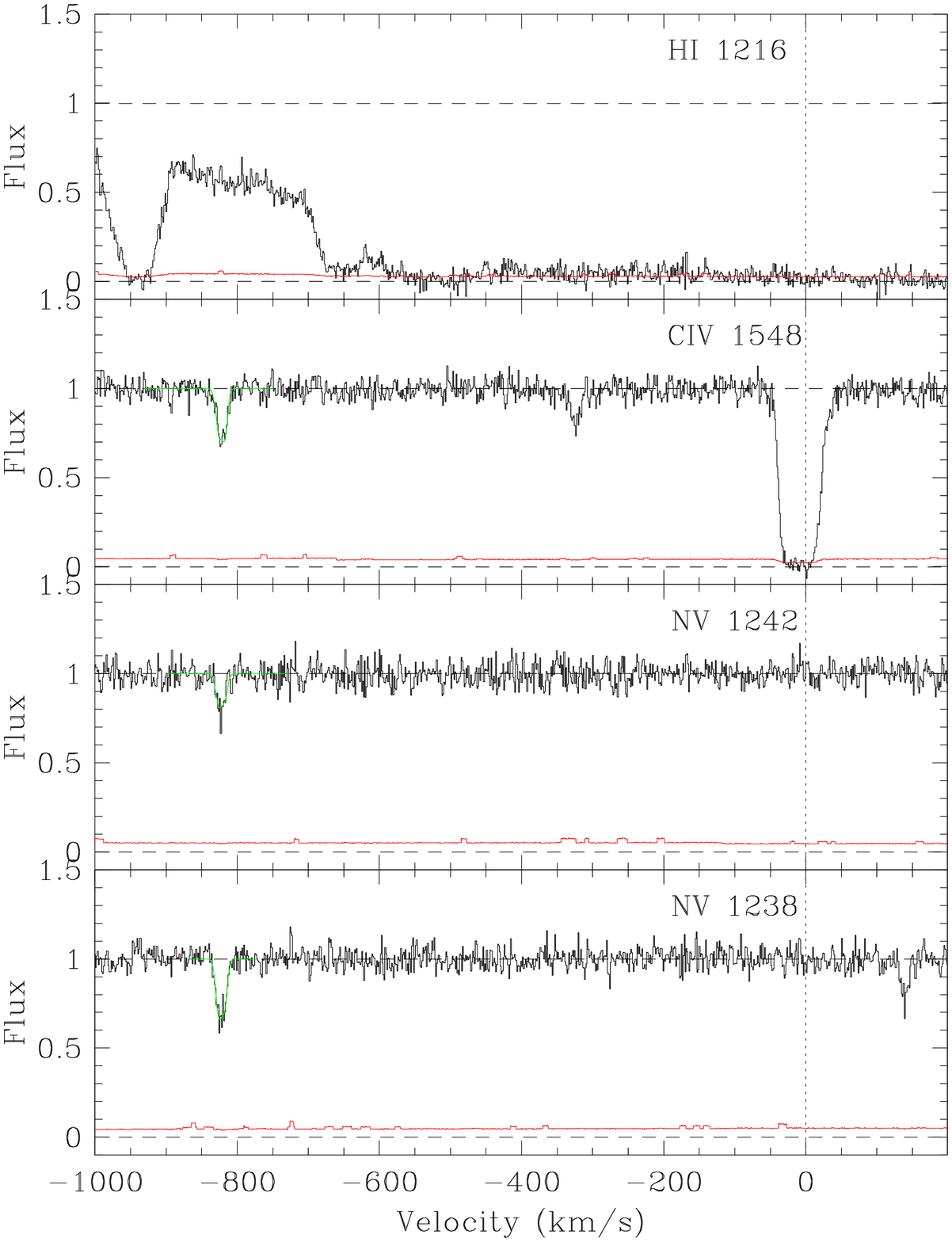}}}}
\caption{\label{0151_metals_high} Highly offset NV and CIV in the PDLA
towards Q0151+048.  The lower solid (red) line shows the error
array.  When fits have been derived using VPFIT, those fits are overlaid in
green.  Velocities are plotted relative to $z_{\rm abs}$=
1.9342. }
\end{figure}

Selected metal lines for the PDLA towards Q0151+048 are shown in
Figures \ref{0151_metals} and \ref{0151_metals_high}.  As described by
M\o ller et al (1998) and Fynbo et al. (1999, 2000), Q0151+048 has a
close, but fainter, companion, Q0151+048B, which is separated by 3.27
arcsec ($\sim 27.5$ \hkpc).  M\o ller et al. (1998) give the redshift
of Q0151+048 to be $z_{\rm em} = 1.922 \pm 0.003$ and a slightly
higher value for Q0151+048B of $z_{\rm em} = 1.937 \pm 0.005$.  As
shown by Zafar et al. (in preparation) Q0151+048B shows no DLA
absorption, despite its relatively small transverse separation.  Using
the same technique described in Section \ref{z_sec}, we re-evaluate
the redshift of Q0151+048 and Q0151+048B from X-shooter spectra kindly
provided by Johan Fynbo and Tayyaba Zafar. We determine $z_{\rm em}$ =
1.9225$\pm 0.003$ for Q0151+048 and $z_{\rm em}$ = 1.9237$\pm 0.003$
for Q0151+048B, where both redshifts are based on the Mg II line.
However, we caution that the spectrum of Q0151+048B is affected by a
lot of structure in the continuum shape which may compromise the
accuracy of the redshift (the quoted error only accounts for known
systematic offsets of emission lines from the systemic value).
Interestingly, the PDLA studied by Rix et al. (2007) towards
Q2343$-$415 also has a nearby QSO companion (Q2343+125) separated by
680 \hkpc\ and $-$100 \kms.

N(FeII) is derived from a simultaneous \textsc{VPFIT} of FeII $\lambda$ 1608
and FeII $\lambda$ 1144.  Similarly, N(SiII) is determined by
simultaneously fitting the lines at $\lambda$ 1304 and 1526 \AA.  The
NI column density was determined from a simultaneous fit of NI
$\lambda$ 1199 and NI $\lambda$ 1200.2.  CII$^{\star}$ is detected,
but its low optical depth precludes an accurate Voigt profile fit; its
column density is derived from the AODM.  By combining
N(CII$^{\star}$) with \nhi\ it is possible to determine the cooling
rate, $l_c$, which Wolfe et al. (2008) have recently shown to be
bi-modal.  We calculate a cooling rate log $l_c = -26.86$ which
qualifies this PDLA as a `high cool' (i.e. high cooling rate) system.
Such systems typically have higher metallicities, dust-to-gas ratios
and velocity spreads than the low cooling rate DLAs.  However, the
PDLA towards Q0151+048 has a relatively narrow velocity structure and
apparently low metallicity ([S/H]$\le-2.03$).  Either this PDLA is
rather unusual amongst the `high cool' population, or the column
densities may be affected by ionization.  However, it should be noted that
it lies towards the lower end of the $l_c$ distribution for high cool
systems.  The low \nhi\ of this system
make the latter possibility a likely explanation and we present more
evidence for ionization in low \nhi\ PDLAs in Section \ref{ion_sec}.

Both CIV and SiIV are very strong and may be partly saturated.  The
velocity structure is simple and coincides with that of the low ions,
see Figure \ref{0151_metals}, which is usually not the case for
intervening DLAs. AODM and \textsc{VPFIT} give consistent lower limits for CIV.
N(SiIV) is derived from the $\lambda$ 1402 line due to mild saturation
in the bluer member of the doublet.  In Section \ref{ion_sec} we
discuss the relative contributions of SiII and SiIV in this PDLA
and the implications of coincident velocity structure.

Although no NV is found at the redshift of the main absorber, there is
NV offset by $\sim -825$ \kms.  This velocity corresponds to a
redshift that is well offset into the \lya\ wing, see Figure
\ref{0151_metals_high}, and does not appear to be associated with high
column density HI.  The NV is well-fit in VPFIT (Figure
\ref{0151_metals_high}, so there is no evidence for partial coverage
in the current data.  The NV is accompanied by weak CIV, but no SiIV
(OVI is not covered by the spectra).
The column densities of this component are not included in the main
table (Table \ref{abund_table}) as this absorption does not appear to
be associated with the same PDLA system.  From \textsc{VPFIT} we
determine N(NV)=13.12$\pm$0.03 from the $\lambda \lambda$ 1238, 1242
doublet and N(CIV)=12.89$\pm$0.02.  This offset absorption is
qualitatively similar to the intervening systems studied by Schaye et
al. (2007), who describe a population of absorbers with significant
column densities of CIV and NV, but with little HI.  Weidinger et
al. (2005) also find orphaned NV near a proximate Lyman limits system
with even larger velocity offset.  Fechner \& Richter (2009) give the
line of sight number density of intervening NV absorbers to be
dN/dz=1.16.  For a $\Delta z = \pm 0.01 = 0.02$ (which corresponds to
$\pm \sim$ 1000 \kms\ at $z=2$) there is a 0.2\% random chance that an
intervening NV system lies this close to a PDLA.  We discuss the
offset NV further in Section \ref{discussion_sec}.

\subsection{J1131+6044, log \nhi\ = 20.50}

\begin{figure}
\centerline{\rotatebox{0}{\resizebox{9cm}{!}
{\includegraphics{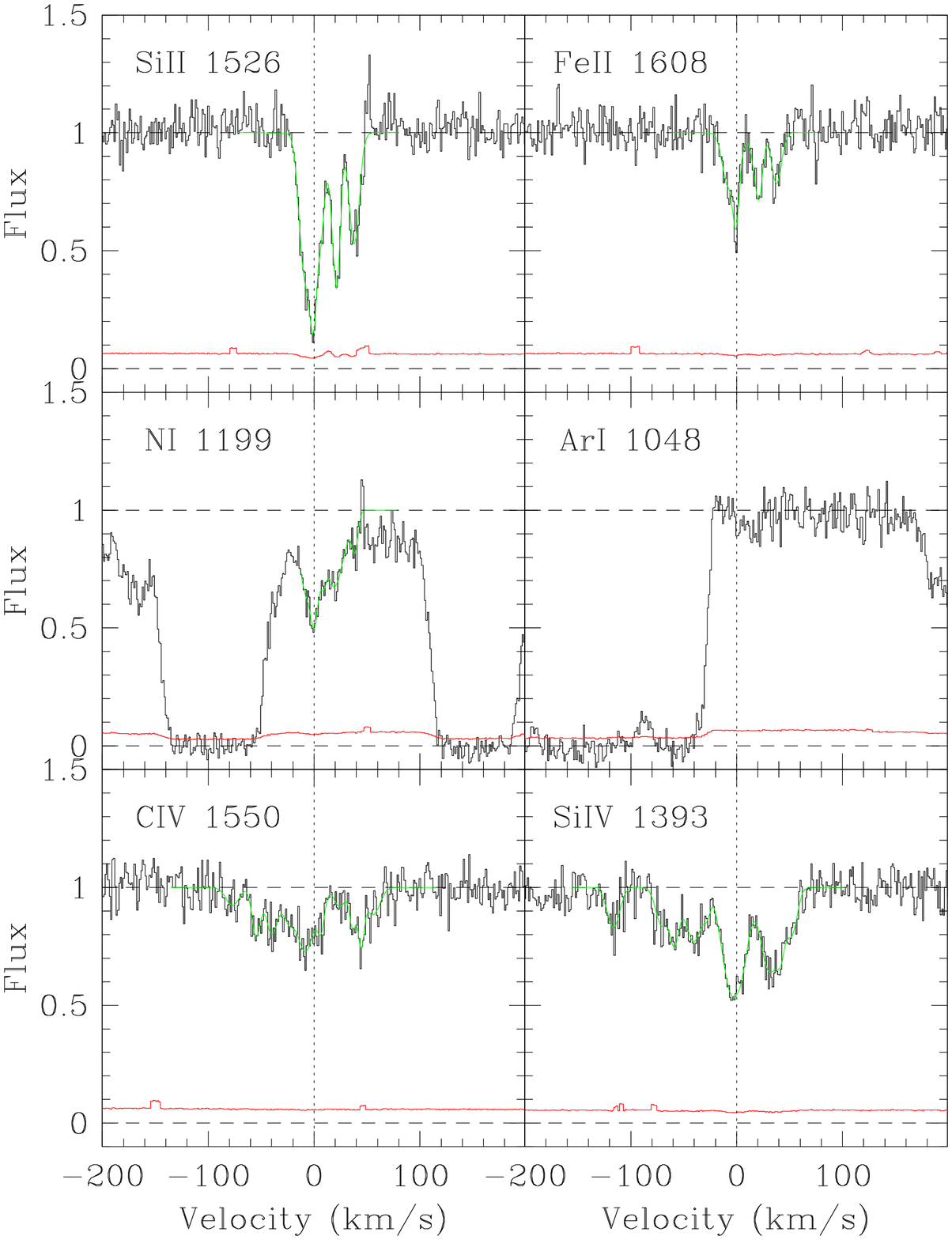}}}}
\caption{\label{1131_metals} Selected metal line transitions in the
PDLA towards J1131+6044.  The lower solid (red) line shows the error
array.  $b$-values for this absorber range from 2 -- 8 \kms\ for the
low ionization species and 2 -- 10 \kms\ for the higher ionization
species.  When fits have been derived using VPFIT, those fits are
overlaid in green.  Velocities are plotted relative to $z_{\rm abs}$=
2.87562.}
\end{figure}

\begin{figure}
\centerline{\rotatebox{0}{\resizebox{9cm}{!}
{\includegraphics{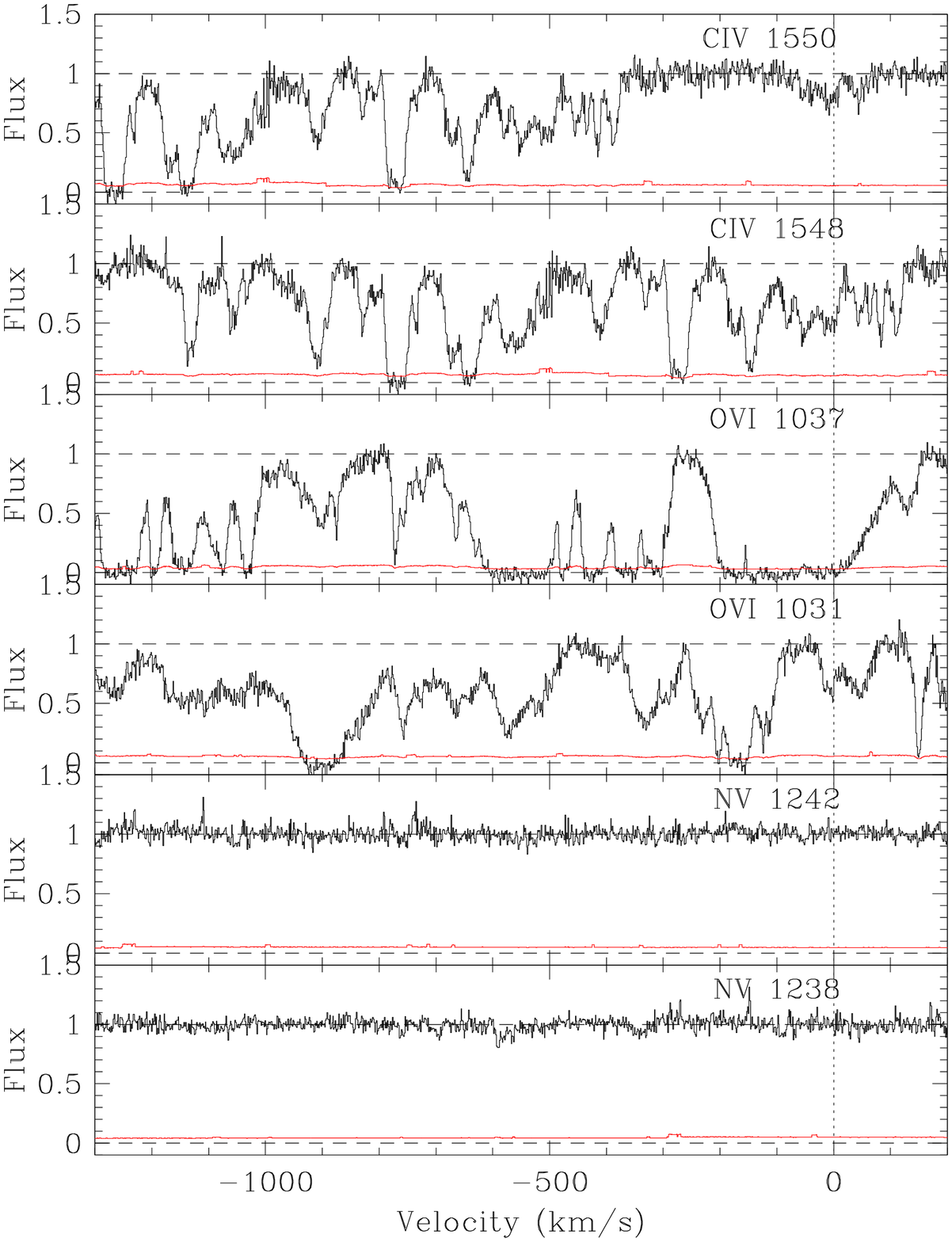}}}}
\caption{\label{1131_metals_high} CIV, OVI and NV coverage in the PDLA
towards J1131+6044 over an extended velocity scale.  The lower solid
(red) line shows the error array.  OVI is detected at $v \sim -750$ \kms\ and
CIV is present from $-1150$ to approximately $-500$ \kms.  There is no
definitive detection of NV.  Velocities are plotted relative to
$z_{\rm abs}$= 2.87562.}
\end{figure}

Selected metal line transitions for the PDLA towards J1131+6044 are
shown in Figure \ref{1131_metals}.  The SiII column density is
determined solely from the $\lambda$ 1526 line, due to saturation of
SiII $\lambda$ 1260, blending of the SiII $\lambda$ 1304 line and no
coverage of SiII $\lambda$ 1808.  Similarly, the only detected FeII
line is at $\lambda_0$=1608 \AA.  Fixing the velocity structure to
match that of the SiII fit produces an excellent fit.  Of the NI
triplet, only the bluest transition at $\lambda_0$=1199 \AA\ is
unblended.  We determine N(NI) using both the fixed component model
derived from SiII and using a free fit.  The former produces a
relatively poor fit (but does not indicate \lya\ blending) to the data
and yields N(NI)=13.87$\pm$0.09.  The latter yields a slightly lower
N(NI)=13.76$\pm$0.08.  We adopt an intermediate value of
13.8$\pm$0.15.  There is a weak (5 $\sigma$) feature at the expected
position of ArI 1048 whose column density can be determined from the
AODM.  Ideally, detection of ArI $\lambda$ is required to confirm that
this is not contamination by weak \lya.  However, since ArI $\lambda$
1066 is not detected to confirm the identification, we take the
conservative approach and use the AODM-derived column density as a
limit.

Although this PDLA has a relatively high $\Delta V$ ($\sim$ 2400 \kms)
it still demonstrates some interesting properties, most notably in its
high ionization lines. At $v \sim 0$ the CIV and SiIV absorption is
relatively weak.  The SiIV $\lambda \lambda$ 1393, 1402 doublet is
fitted simultneously.  The CIV $\lambda$1548 line is blended with
extended negative velocity gas (see below and Figure
\ref{1131_metals_high}) so N(CIV) is determined solely from CIV
$\lambda$ 1550.  There is no NV at $v=0$ but we are not able to
determine whether or not OVI is present due to the blending in at
least one of the doublet components.  However, the kinematic
similarity between absorption at the expected wavelength of
OVI$\lambda$ 1031 and CIV $\lambda$ 1550 is suggestive (Figure
\ref{1131_metals_high}).  We do find highly offset absorption of
strong CIV and OVI (but no NV) at $v \sim -750$ \kms.  Since we can not
estimate the total OVI column density due to blending, no value
is given in Table \ref{highion_table}.  The CIV
extends\footnote{There are two weakly detected features in this
approximate velocity range that could be NV $\lambda$1238.  However,
NV $\lambda$ 1242 is not detected at these extended velocites, so
these identifications can not be confirmed with the present data.}
from approximately $-500$ to $-1150$ \kms.  We can not confirm the
full extent of the OVI in velocity space due to various \lya\ blends, see
Figure \ref{1131_metals_high}.  The metal column densities in this
extended component are not included in the values in Table
\ref{highion_table}.

\subsection{J1240+1455, log \nhi\ = 21.3}\label{1240_sec}

\begin{figure*}
\centerline{\rotatebox{270}{\resizebox{12cm}{!}
{\includegraphics{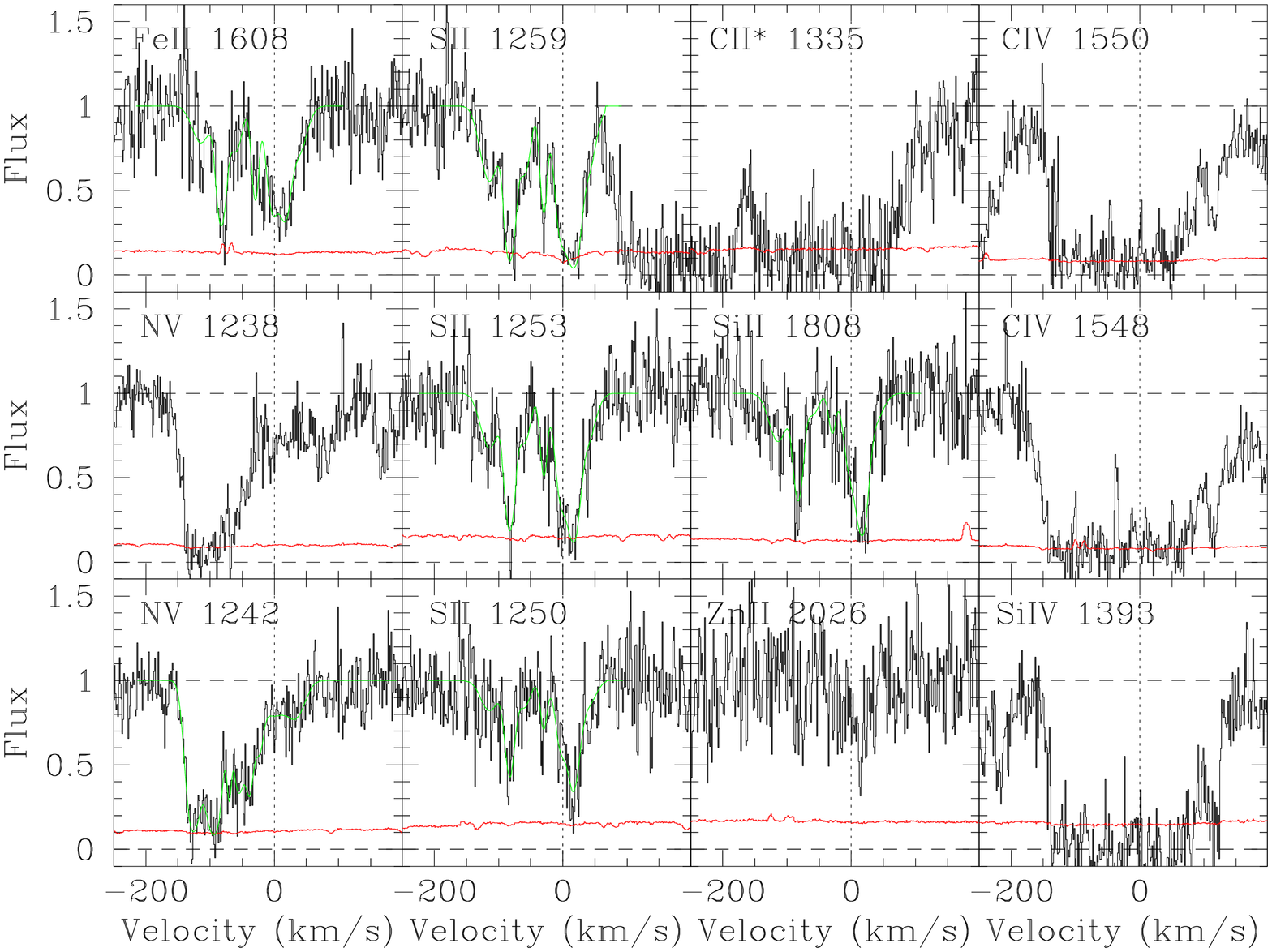}}}}
\caption{\label{1240_metals} Selected metal line transitions in the PDLA
towards J1240+1455.  The lower solid (red) line shows the error
array.  When fits have been derived using VPFIT, those fits are overlaid in
green.   $b$-values for this absorber range from 7 -- 18
\kms\ for the low ionization species and 6 -- 30 \kms\ for the higher
ionization species.   Velocities are plotted relative to $z_{\rm abs}$=
3.1078.}
\end{figure*}

Selected metal lines are shown for the PDLA towards J1240+1455
in Figure \ref{1240_metals}.
This PDLA was studied by Hennawi et al. (2009) using the original SDSS
spectrum.  Like several other PDLAs (M\o ller \& Warren 1993; M\o ller et
al. 1998; Leibundgut \& Robertson 1999; Ellison et al. 2002) this
absorber has Lya emission superimposed in the DLA trough.  After
extensive modelling, Hennawi et.  (2009) concluded that the \lya\
emission, is likely to be associated with the QSO, and not the PDLA
itself.  

This is a relatively faint QSO and the HIRES spectrum exhibits a low
S/N.  Nonetheless, we are able to detect a number of metal lines due
to the high \nhi\ (see Figure \ref{1240_metals}).  For example, all
three lines in the SII $\lambda \lambda \lambda$ 1250, 1253, 1259
triplet are detected and are fitted simultaneously.  We check for
saturation by additionally calculating the AODM column densities,
which yield the same column density in each triplet. Due to the low
S/N, we determine N(FeII, SiII) in three different ways: using a fixed
component \textsc{VPFIT} model based on the component structure of
SII, a model where the component structure can vary and finally the
AODM.  All give very consistent answers, so we adopt that of the fixed
component \textsc{VPFIT} model.  Since the optical depths in the line
centres of the SiII and FeII lines are less than in the demonstrably
unsaturated SII lines, saturation does not seem to be an issue.  The
Fe and Cr abundances are both low in this system (e.g.  relative to
Zn), indicating that dust depletion may be significant.  Strong, broad
NV absorption is detected, offset by $\sim -$120 \kms\ from the
strongest low ion component.  Only the weaker NV $\lambda$ 1242 line
is fitted, although even this may be partly affected by saturation.
We therefore conservatively quote the fit as a lower limit in Table
\ref{highion_table}.  CIV and SiIV are both heavily saturated and
CII$^{\star}$ is blended.

\subsection{J1604+3951, log \nhi\ = 21.75}\label{1604_sec}

\begin{figure*}
\centerline{\rotatebox{270}{\resizebox{12cm}{!}
{\includegraphics{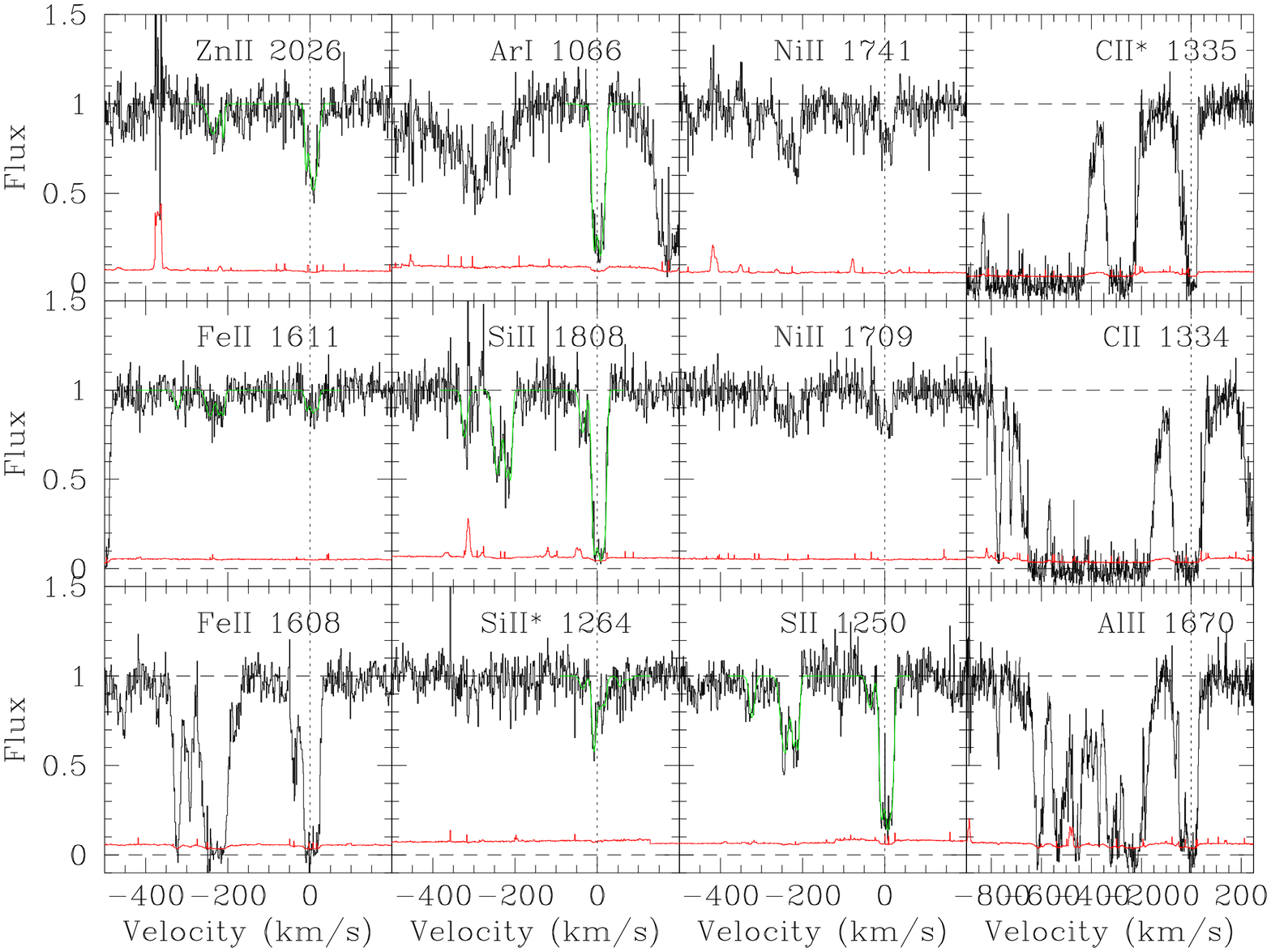}}}}
\caption{\label{1604_low} Low ionization metal line transitions in the
PDLA towards J1604+3951.  The lower solid (red) line shows the error
array.  When fits have been derived using VPFIT, those fits are
overlaid in green.  $b$-values for this absorber range from 3 -- 12
\kms\ for the low ionization species. Velocities are
plotted relative to $z_{\rm abs}$= 3.1670. Note the different velocity
scale of the last column.  The CII$^{\star}$ line is blended to the
blue with CII. }
\end{figure*}

\begin{figure*}
\centerline{\rotatebox{270}{\resizebox{12cm}{!}
{\includegraphics{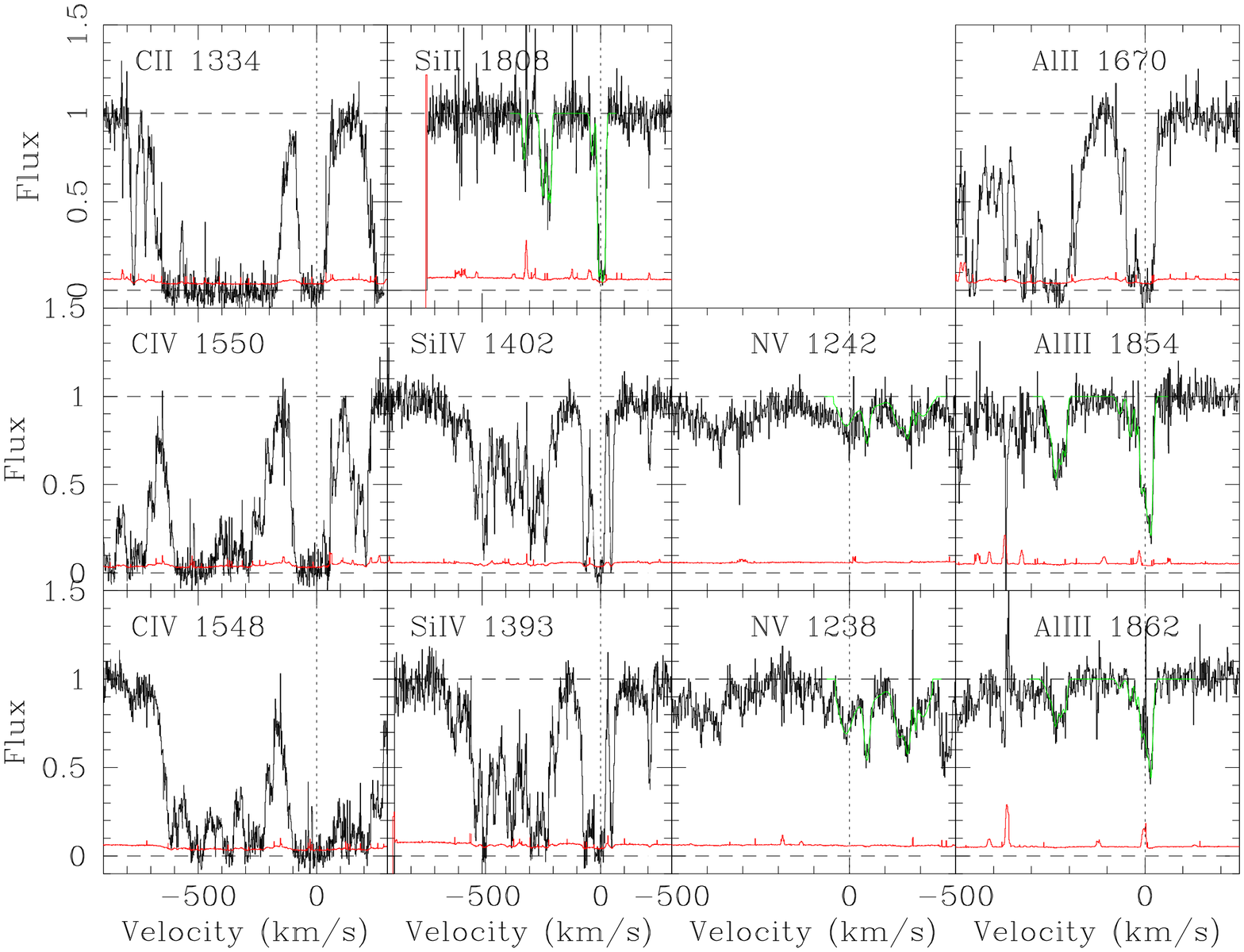}}}}
\caption{\label{1604_high} High ionization metal line transitions in
the PDLA towards J1604+3951.  The lower solid (red) line shows the
error array.  When fits have been derived using VPFIT, those fits are
overlaid in green.  $b$-values for this absorber range from 3 -- 25
\kms\ for the high ionization species.  Velocities are plotted relative
to $z_{\rm abs}$= 3.1670. The red component of CIV 1548 is blended
with the blue component of CIV 1550.  The upper row shows some of the
low ions (repeated from Figure \ref{1604_low}) for comparison.}
\end{figure*}

The PDLA towards J1604+3951 exhibits two main components in the low
ionization gas separated by about 250 \kms, see Figure \ref{1604_low}.
However, the strongest transitions, such as CII $\lambda$ 1334 and
AlII $\lambda$ 1670 show almost continuous absorption over 600 \kms.
Singly and multiply ionized species exhibit quite similar velocity
structure, e.g. SiII and SiIV (Figure \ref{1604_high}).  In the alpha
elements (Si and S) the redder component is stronger, but in Fe-peak
(Fe and Ni) the blue component is stronger.  The absorption structure
of ZnII is more akin to Si and S than Fe and Ni, even though it is
often considered to track the Fe-peak.  We return to this point below.
 
SII $\lambda$ 1250 and $\lambda$ 1253 are fitted simultaneously and we
apply the same model for SiII $\lambda$ 1808.  FeII $\lambda$1608 is
saturated in several components and yields a lower limit of
N(FeII)$>$15.15.  Fortunately, FeII $\lambda$1611 is detected, albeit
with low optical depth.  We try both a fixed (tied to the SII
structure) and free velocity model; both give a consistent answer of
N(FeII)=15.35$\pm$0.03.  The AODM gives a slightly higher value of
N(FeII)=15.45$\pm$0.2.  We adopt an intermediate value of
N(FeII)=15.4$\pm$0.15.  CII$^{\star} \lambda$1335 is saturated at
$v\sim0$ \kms\ and the negative velocity component is blended with CII
$\lambda$1334.  We therefore quote a lower limit that is very
conservative since it does not (cannot) account for the negative
velocity gas.  Saturation and blending mean that the lower limit for
CII $\lambda$ 1334 is similarly conservative.  As for FeII, we attempt
three different approaches for Zn, all give consistent values, so an
average is taken.  ArI $\lambda$ 1048 is blended, as are the weaker
components of  ArI $\lambda$ 1066.  However, the strongest velocity
component of  ArI $\lambda$ 1066 appears clean.  Whilst it is therefore
not possible to report a total N(ArI), we fit the  ArI $\lambda$ 1066
line with the same velocity structure as SiII to derive N(ArI)=14.45
for the $v=0$ complex.  The SiII column density for these components
is 15.96, yielding [Ar/Si]=$-0.40$.

Figure \ref{1604_high} shows absorption from the more highly ionized
species.  CIV and SiIV are particularly strong. Due to the wide
velocity structure, CIV is self-blended.  The limit for N(CIV) is
therefore derived from the blue (negative velocity) component of CIV
$\lambda$1548.  NV is detected, but the poor continuum determination
makes fitting Voigt profiles challenging. A simultaneous fit to NV
$\lambda$1238 plus NV $\lambda$ 1242 from $-100$ to $250$ \kms\ yields
N(NV)=14.14$\pm$0.02 (the error does not include errors in the
continuum).  There appears to be further NV at $-400$ \kms, where CIV
and SiIV are also seen, but not the low ionization species.  Fits to
this bluer component are hampered by the poor continuum placement and
possible blending from other lines and not accounted for in the column
density quoted in Table \ref{highion_table}.

Considering the relative abundances produces some clues to the
difference in the velocity structure of the low ions.  The column
densities for the two main velocity components at $-240$ and 0 \kms\
are given in Table \ref{J1604_table}.  Also tabulated are the relative
abundances of some key elemental ratios.  Both components have [S/Si]
ratios consistent with the solar value, as expected from studies of
Galactic stars (Chen et al. 2002).  Similarly, the [S/Zn] ratio is
consistent with mild $\alpha$ element enhancement in both components
(where S and Zn are both undepleted and therefore a useful combination
for this assessment, e.g. Nissen et al. 2004).  However, whereas the
blue component has a solar [Zn/Fe] ratio, the gas at $v \sim 0$ \kms\
has a very high value: [Zn/Fe]=0.75.  These results are strongly
indicative of depletion patterns varying within the galaxy.

There is a possible detection of SiII$^{\star} \lambda$ 1264.
Although excited transitions of Si and Fe have been previously
reported in DLAs associated with GRBs (e.g. Prochaska, Chen \& Bloom
2006; Vreeswijk et al. 2007), this is the first possible detection in
a QSO (P)DLA.  The fit gives N(SiII$^{\star}$)=12.6$\pm$0.1.  Although
this column density is 16 times larger than the 3$\sigma$ upper limit
derived for the absorber PKS 1443+27 (Howk et al. 2005), the \nhi\ of
the PDLA is also almost 10 times higher.  Unfortunately, the
saturation of the CII$^{\star}$ transition precludes an estimate of
the warm neutral medium fraction in this absorber.

\subsection{J2321$+$1421, log \nhi\ = 20.70}\label{2321_sec}

\begin{figure*}
\centerline{\rotatebox{270}{\resizebox{12cm}{!}
{\includegraphics{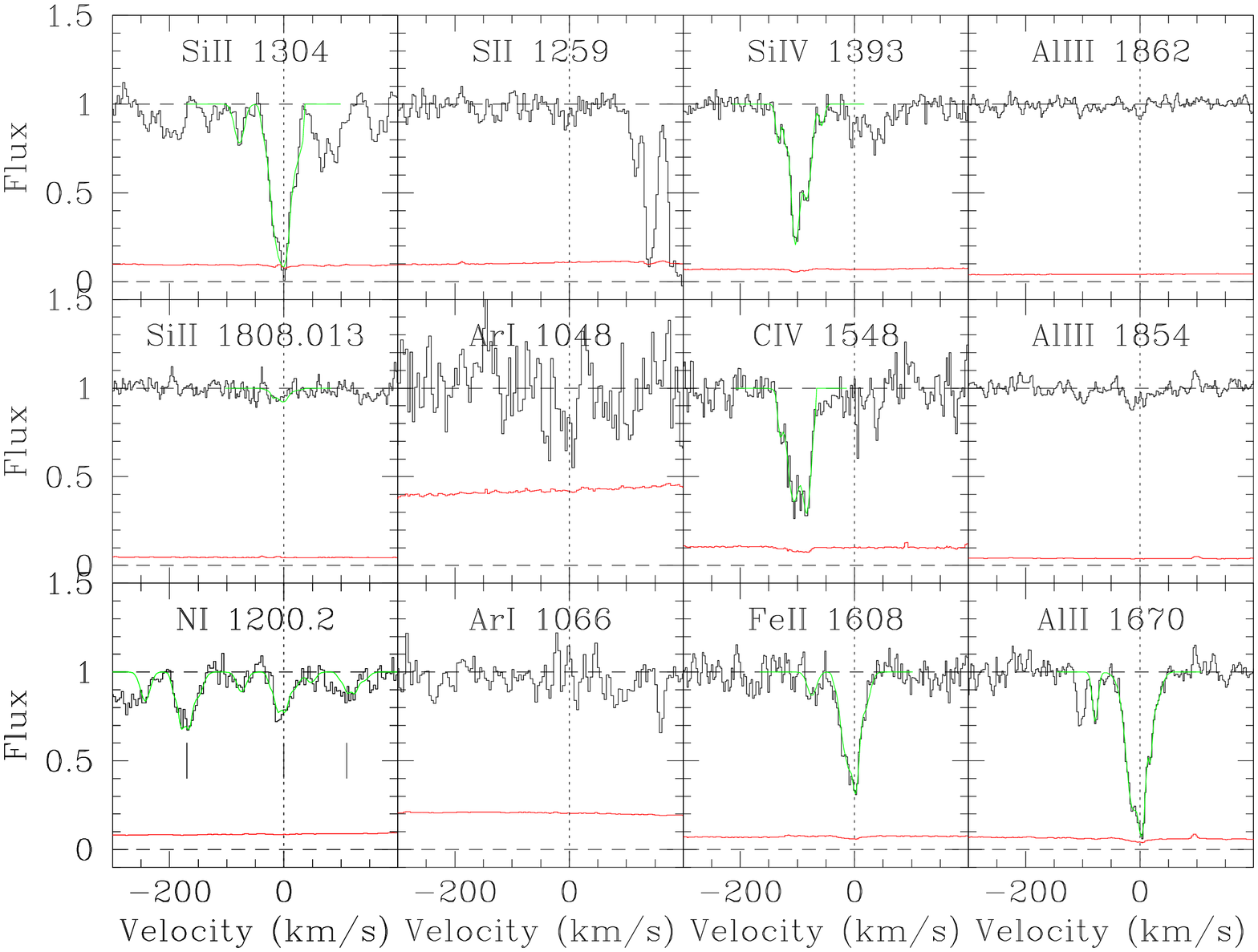}}}}
\caption{\label{2321_metals} Selected metal line transitions in the PDLA
towards J2321$+$1421.  The lower solid (red) line shows the error
array.  When fits have been derived using VPFIT, those fits are overlaid in
green.   $b$-values for this absorber range from 2 -- 12
\kms\ for the low ionization species and 5 -- 25 \kms\ for the higher
ionization species.   Velocities are plotted relative to $z_{\rm abs}$=
2.5731. The vertical ticks in the lower left panel show the strongest
components in each of the NI triplet lines.}
\end{figure*}

The PDLA towards J2321+1421 has a relatively simple velocity structure
with most of the absorption centred within $\pm$40 \kms\ of $z_{\rm
abs} = 2.5731$, see Figure \ref{2321_metals}. N(FeII) is determined
from a simultaneous fit to FeII $\lambda$ 1608 and FeII $\lambda$
2374, both of which are unsaturated.  SiII $\lambda$ 1808 is a maginal
detection, whereas SiII $\lambda$ 1304 and SiII $\lambda$ 1526 are
mildly saturated.  Nonetheless, combining the 3 transitions we are
able to constrain the total SiII column density.  The region of
spectrum around the strongest CrII line at $\lambda_0$=2056 \AA\ is
relatively noisy, so we derive an upper limit from the non-detection
of CrII $\lambda$ 2062.  There is minor blending of NI with \lya\
forest lines but simultaneous fitting of the triplet gives a
consistent fit.  There is weak (6 $\sigma$) absorption at
approximately the expected position of AlIII $\lambda$ 1854;
N(AlIII)=12.2 is determined from the AODM.  However, this is
inconsistent with the non-detection of AlIII 1862 for which we derive
a 3 $\sigma$ limit of N(AlIII)$<$11.9.  Given the weakness of the 1854
\AA\ line and the possibility of contamination we give 12.2 as a
conservative upper limit.  CIV and SiIV are detected, but offset by
$-100$ \kms\ from the strongest low ion absorption, but coincident
with much weaker column density components e.g. in AlII and FeII (see
Figure \ref{2321_metals}).  The spectral region around the ArI doublet
is fairly noisy.  ArI $\lambda$ 1066 is not detected, but there is a 5
$\sigma$ feature at the expected velocity of ArI $\lambda$ 1048.
However, the S/N at this wavelength is only around 2 per pixel, so we
consider this unreliable and thterefore conservatively adopt the upper
limit from ArI $\lambda$ 1066.

\begin{center}
\begin{table*}
\begin{tabular}{lcccccccccc}
\hline
QSO      &     log N(FeII)      & log N(CII)     & log N(SiII)    & log N(SII)     & log N(OI)      & log N(ZnII)    & log N(CrII)    & log N(NI)     & log N(AlII) & N(ArI) \\ \hline 
J0140$-$0839    & $<$12.73       & 14.13$\pm$0.08 & 13.51$\pm$0.09 & $<$13.33       & 14.69$\pm$0.01 & blend          & $<$12.39      & $<$12.38      & 11.82$\pm$0.04 & $<$12.82 \\
J0142$+$0023   & 13.7$\pm$0.1   &    ...         & 14.15$\pm$0.03 & 13.28$\pm$0.06 & $>$15.01       & $<$11.50       & $<$12.17       &      blend    & 12.73$\pm$0.01 & $<$12.57\\
Q0151+048     & 13.70$\pm$0.01 & $>$14.43       & 14.01$\pm$0.05 & $<$13.47       & $>$14.84       & $<$11.81       & $<$12.45       & 13.06$\pm$0.05& 12.57$\pm$0.05 & ... \\
J1131+6044     & 13.76$\pm$0.03 & $>$14.55       & 14.49$\pm$0.13 & $<13.29$       & $>$14.82       &  ...           & ...            & 13.8$\pm$0.15 & ... & $<$12.52\\
J1240+1455     & 14.60$\pm$0.03 & sat/blend      & 15.93$\pm$0.03 & 15.56$\pm$0.02 & $>$15.24       & 12.90$\pm$0.07 & $<$13.02       &      blend    & $>$13.56 & ...\\
J1604+3951     & 15.40$\pm$0.15 & $>$15.28       & 16.09$\pm$0.02 & 15.70$\pm$0.02 & blend          & 13.0$\pm$0.1   & ...            &  ...          & $>$14.00 & yes/blend\\
J2321+1421     & 14.18$\pm$0.03 & $>$14.68       & 14.45$\pm$0.04 & $<$13.60       & $>$15.10       & $<$11.84       & $<$12.57      & 13.64$\pm$0.03& 12.99$\pm$0.02 &$<$13.33 \\  
\hline 
\end{tabular}
\caption{\label{lowion_table}Column densities (in \cm) of low ionization species.}
\end{table*}
\end{center}

\begin{center}
\begin{table*}
\begin{tabular}{lcccccc}
\hline
QSO &          log N(CIV) & log N(SiIV) & log N(NV) & log N(OVI) & log N(AlIII) & N(CII$^{\star}$)\\
\hline 
J0140$-$0839  & $<$12.18 & $<$12.20 & $<$12.96 & blend & $<$ 11.52 & $<12.41$ \\
J0142$+$0023  & 14.25$\pm$0.01 & 13.73$\pm$0.01 & $<$12.29 & blend & $<12.4$ & ... \\
Q0151+048     & $>$14.50  &  13.75$\pm$0.01  & $<$12.66 & ... & 12.3$\pm$0.1 & 13.0$\pm$0.2 \\
J1131+6044 &    13.85$\pm$0.05 & 13.33$\pm$0.02  &  $<$12.68 & yes/blend &  ...     & $<12.51$ \\                   
J1240+1455    & $>$15.13 & $>$14.31 & $>$14.86 & ... & blend & blend \\
J1604+3951    & $>$15.05 & $>$14.74 & 14.14$\pm$0.02 & blend & 13.44$\pm$0.02 & $>$14.30\\
J2321+1421   & 13.81$\pm$0.05 & 13.37$\pm$0.01 & $<$12.62 & blend & $<$12.2 & $<12.55$\\

\hline 
\end{tabular}
\caption{\label{highion_table}Column densities  (in \cm) of high ionization and excited species}
\end{table*}
\end{center}

\begin{center}
\begin{table}
\begin{tabular}{lcc}
\hline
  & Blue cpt.  & Red cpt. \\
  & $v\sim-240 $\kms & $v \sim 0$ \kms\\
\hline 
N(ZnII) & 12.43 & 12.88\\
N(FeII) & 15.27 & 14.97 \\
N(SiII) & 15.50 & 15.96 \\
N(SII) & 15.16 & 15.55 \\
 & & \\
$[$Zn/Fe$]$  &  0.00  &  +0.75 \\
$[$S/Si$]$ & 0.01 & $-0.06$\\
$[$S/Zn$]$ & 0.18 & 0.12 \\
\hline 
\end{tabular}
\caption{\label{J1604_table}Column densities (\cm) and abundance
ratios in two components in PDLA J1604+3951}
\end{table}
\end{center}

\begin{center}
\begin{table*}
\begin{tabular}{lcccccccccc}
\hline
QSO    & log N(HI)               & [C/H]  &  [N/H] &   [O/H]    & [Si/H]  & [S/H]    &  [Fe/H]  & [Zn/H]   & [Cr/H] & [Ar/H] \\
\hline 
J0140$-$0839  & 20.75$\pm$0.05 & $-3.01$ & $<-4.15$ & $-2.72$ & $-2.75$ & $<-2.58$ & $<-3.47$ &     ...  & $<-1.99$ & $<-2.33$ \\
J0142$+$0023  & 20.38$\pm$0.05 & ...     & ....     & $>-2.03$& $-1.74$ & $-2.26$ & $-2.13$ & $<-1.49$  & $<-1.84$ & $<-2.21$\\
Q0151+048     & 20.34$\pm$0.02 & $>-2.30$ & $-3.06$ & $>-2.16$ & $-1.84$ & $<-2.03$ & $-2.09$ & $<-1.14$ &  $<-1.52$ & ... \\
J1131+6044    & 20.50$\pm$0.15 & $>-2.34$ & $-2.48$ & $>-2.34$ & $-1.52$ & $<-2.37$ & $-2.19$ & ... & ... & $<-2.38$\\
J1240+1455    & 21.3$\pm$0.2 & ...     & ...     & $>-2.72$ & $-0.88$ & $-0.90$  & $-2.15$  & $-1.01$ & $<-1.91$ & ... \\
J1604+3951   & 21.75$\pm$0.2 & $>-2.86$ & ...     & ...      & $-1.17$ & $-1.21$  & $-1.80$ & $-1.36$ & ... & ...\\
J2321+1421   & 20.70$\pm$0.05 & $>-2.41$ & $-2.84$ & $>-$2.26 & $-1.76$ & $<-2.26$ & $-1.97$ & $<-1.47$ & $<-1.76$ & $<-1.77$\\
\hline 
\end{tabular}
\caption{\label{abund_table}Abundances relative to the solar scale of
Asplund et al. (2005), except for argon for which we use the value
given in Asplund et al (2009).  The approximation of  N(X)= N(XII) or N(XI) 
has been assumed.}
\end{table*}
\end{center}

\section{Additional proximate and intervening DLAs from the 
literature}\label{lit_sec}

Although the study presented here represents the first systematic
study of a sample of PDLAs, a small number of proximate systems
have been included in literature studies of intervening absorbers.
In this section, we search the literature for PDLAs that meet our
$\Delta V$ selection criterion and with measured metal column densities
that can be used to enlarge our sample.  We also describe the
compilation of a comparison sample of intervening DLAs with which the
PDLAs can be compared.

\subsection{PDLAs in the literature}

We searched for additional PDLAs with known abundances by trawling the
catalogue of Dessauges-Zavadsky et al. (in prep.).  Emission redshifts
were taken from the references given in Dessauges-Zavadsky et al. (in
prep); if no emission redshift was present in the referenced paper, we
used SIMBAD.  Due to the numerous different techniques and inherent
uncertainties in emission redshift determination, a first cut was made
for literature DLAs with $\Delta V < 5000$ \kms, of which there are 15
in the Dessauges-Zavadsky catalogue.  We obtained literature spectra
for 14/15 of these targets that were suitable for a
re-assessment\footnote{We were unable to locate a suitable spectrum
for the re-measurement of the emission redshift of Q0201+365.} of
$z_{\rm em}$ using the technique described in Section \ref{z_sec}. In
this way, all of the PDLA emission redshifts and hence the $\Delta V$
values are computed in a consistent way.  In Table \ref{lit_z_table}
we list the PDLA candidates selected from the literature together with
updated emission redshifts and re-computed $\Delta V$ where available.
There is no obvious trend of $\Delta V_{\rm new} - \Delta V_{\rm old}$
with redshift and positive and negative changes are present in 
approximately equal number.  Only one QSO's redshift changes by $>$
1000 \kms, name Q0425$-$5214 (CTS 436) whose redshift is obtained from
the \lya\ line (Maza et al. 1995).  Not only is this line a poor
indicator of systemic redshift, but the spectrum is of very low
disperison (30\AA\ resolution) are reported as being accurate to only
$\pm$0.02.
We discard absorbers with $\Delta V >$ 3000 \kms.  To complement the
abundances determined for our new data sample (Section
\ref{coldens_sec}) column densities for literature PDLAs with $\Delta
V <$ 3000 \kms\ are taken from Dessauges-Zavadsky et al. (in prep.)
and combined with our new echelle sample.  The final PDLA sample
therefore contains 7+9 (new plus literature respectively) PDLAs.

In addition to the column densities available in the literature
(Table \ref{lit_abund_table}) we determine the column density
of SII  towards J2340$-$00 from the extant HIRES spectrum.
All 3 transitions in the SII triplet at $\lambda \lambda \lambda$
1250, 1253, 1259 \AA\ are well detected.  We determine the 
column densities using the AODM and find that all 3 lines yield
values in excellent agreement.

\begin{center}
\begin{table*}
\begin{tabular}{lccrccc}
\hline
QSO & $z_{\rm em}$ lit. & $z_{\rm abs}$  & $\Delta V$ lit. (\kms) &  $z_{\rm em}$ new & $\Delta V$ new (\kms)&  Include?\\
\hline
Q1157+014 &  1.990 &  1.944 &  4651 & 1.9920$\pm$0.003 & 4851 & N \\
J2340$-$00 &  2.090 &  2.054 &  3516  &2.0829$\pm$0.003 & 2825 & Y \\
Q2222$-$3939 &  2.18 &  2.154 &  2463  & 2.1832$\pm$0.007 & 2765 & Y\\
Q0425$-$5214 &  2.25 &  2.224 &  2410  & 2.2639$\pm$0.007  & 3690 & N \\
Q0841+12 &  2.500 &  2.476 &  2064  & 2.4934$\pm$0.003 & 1498 & Y \\
B0405$-$331 &  2.570 &  2.569 &    84  &2.5775$\pm$0.006 & 714   & Y \\
Q2343$-$BX415 &  2.57393 &   2.5720 & 162 &  2.5742$\pm$0.0005 &  22 & Y \\
Q0528$-$2505 &  2.779 &  2.812 &  $-$2608  &2.7783$\pm$0.007 & $-$2664 & Y \\
Q1354$-$1046 &  3.007 &  2.967 &  3010  & 3.0112$\pm$0.010 & 3324 & N \\
Q2059$-$360 &  3.09 &  3.083 &   514  &3.0974$\pm$0.009 & 1056 & Y \\
SDSS2100$-$0641 &  3.140 &  3.0924 &  3469 & 3.1295 &2707 & Y \\
J0900+42 &  3.290 &  3.2458 &  3107  &3.2954 & 3484 & N \\
J0255+00  & 3.988 &  3.915  & 4423  &3.9936 & 4759 & N \\
PSSJ0957+33 &  4.227 &  4.178 &  2826  &4.2088 & 1779 & Y \\
\hline 
\end{tabular}
\caption{\label{lit_z_table}Candidate PDLAs taken from the literature.
Redshifts without error bars are taken from Prochaska, Hennawi \& Herbert-Fort
(2008). The final column indicates whether the absorber is included in our
PDLA sample, which requires that the new be $\Delta V < 3000$ \kms.}
\end{table*}
\end{center}

\begin{center}
\begin{table*}
\begin{tabular}{lccccccccc}
\hline
QSO  &  $z_{\rm abs}$ &  log N(HI)  &     log N(FeII)    &     log N(ZnII)    &  log N(SiII)  &    log N(CrII)    &  log N(SII) &  [Si/H] & Reference \\ \hline        
J2340-00     &    2.054   & 20.35$\pm$0.15 & $>$14.97          &  12.63$\pm$0.08 & 15.17$\pm$0.04 &  $<$12.90           &  14.94$\pm$0.05         & $-0.69$ & 12,13 \\
Q2222-3939   &    2.154   & 20.85$\pm$0.10 &  14.42$\pm$0.03 &  $<$11.70            &  14.55$\pm$0.05  & 12.77$\pm$0.04  &  14.08$\pm$0.02 &  $-1.81$ &4,5  \\ 
Q0841+12     &    2.476   & 20.78$\pm$0.08 &  14.50$\pm$0.03 &  11.69$\pm$0.10 & 14.99$\pm$0.03 &   12.89$\pm$0.06  &  14.48$\pm$0.10 & $-1.30$  &6  \\ 
B0405-331    &    2.569   & 20.60$\pm$0.10 &  14.33            &  $<$12.74         &  14.74           &   $<$13.32          &   ...   & $-1.37$  &  1 \\  
Q2343-BX415  &    2.5720  & 20.98$\pm$0.05 &  15.24$\pm$0.02 &  12.90$\pm$0.06 & 15.79$\pm$0.04 &   13.58$\pm$0.04  &  15.38$\pm$0.03 & $-0.70$  &  11 \\  
Q0528-2505   &    2.812   & 21.11$\pm$0.04 &  15.47$\pm$0.02 &  13.27$\pm$0.03 & 16.01$\pm$0.03 &   13.65$\pm$0.12  &  15.56$\pm$0.02 & $-0.61$  &  2,3  \\ 
Q2059-360    &    3.083   & 20.98$\pm$0.08 &  14.52$\pm$0.07 &  ...            & 14.80$\pm$0.05 &   ...             &  14.41$\pm$0.04 & $-1.69$ &  7  \\ 
SDSS2100-0641&    3.0924  & 21.05$\pm$0.15 &  15.36$\pm$0.03 & $<$13.14          & 15.89$\pm$0.02 &   13.69$\pm$0.04  &  15.52$\pm$0.01 & $-0.67$  &  8  \\ 
PSSJ0957+33  &    4.178   & 20.65$\pm$0.15 &  14.13$\pm$0.05 &  ...            & 14.56$\pm$0.01 &   ...             &  14.39$\pm$0.06 & $-1.60$ &  9,10 \\ 
\hline
\end{tabular}
\caption{\label{lit_abund_table}Column densities  (in \cm) for final literature
PDLA sample.  References: 1: Akerman et al. (2005); 2: Lu et
al. (1996); 3: Centurion et al. (2003); 4: Noterdaeme et al. (2008);
5: P. Noterdaeme private communication; 6: Dessauges-Zavadsky et al. (2007);  7
Srianand et al. (2005); 8: Herbert-Fort et al. (2006); 9: Prochaska et
al. (2001); 10: Prochaska et al. (2003b); 11: Rix et al. (2007); 12:
Prochaska et al. (2007b); 13: this work.}
\end{table*}
\end{center}

\subsection{DLA comparison sample}

To compare the PDLAs with intervening DLAs we used the sample of 
Dessauges-Zavadsky
et al. (in prep.).  In order to circumvent the large uncertainties
in $\Delta V$ incurred through $z_{\rm em}$ measurements, we impose
a lower limit of 10,000 \kms\ which yields a sample of 180 intervening 
DLAs with log \nhi\ $\ge$ 20.3 from
the Dessauge-Zavadsky catalogue.  We adopt the column density measurements
compiled by Dessauges-Zavadsky et al., and convert to abundances
using the same solar scale that was applied to the PDLAs (Asplund et
al. 2005).  In Sections \ref{ion_sec} and \ref{abund_sec} we
compare the properties of the PDLAs with the DLAs in order to gain
insight into whether they represent similar populations and
what the effect of QSO proximity is.

\section{Ionization}\label{ion_sec}

The proximity of PDLAs to their background QSO naturally leads to the
question of whether they are strongly affected by the quasar's intense
ionizing radiation.  Thus far, there are conflicting indications in
the literature for studies of individual systems.  Rix et al. (2007),
in their detailed study of a single PDLA towards Q2343$-$BX415,
concluded that ionization corrections were small.  Nonetheless, they
identify absorption from highly ionized species such as NV which are
relatively rarely detected in intervening absorbers and require high
energy photons if produced via photoionization.  One of the PDLAs in
our sample (J1240+1455) has previously been reported to exhibit NV in
the low resolution SDSS spectrum (Hennawi et al. 2009) which we
confirm in our echelle spectra (see Section \ref{1240_sec}).  Fox et
al. (2009) find that $\sim$ 13\% of intervening systems also exhibit
NV and do not find any increase in its incidence in PDLAs with
relative velocities 500$<\Delta V<$ 5000 \kms\ relative to the
intervening systems at $\Delta V>5000$ \kms.  In a general survey of
NV at all \nhi\ column densities, Fechner \& Richter (2009) do find a
proximate excess, a conflict which appears to be linked to the
velocity range used (see discussion below).  Finally, \lya\ emission
in the DLA trough has been found for several PDLAs (M\o ller \& Warren
1993; M\o ller et al. 1998; Leibundgut \& Robertson 1999; Ellison et
al. 2002; Hennawi et al.  2009), a feature which is much rarer amongst
the intervening DLA population.  In this section, we assess various
indications of ionization in the sample of 16 PDLAs, paying special
attention to elements which might indicate ionization by a hard
radiation source, as might be expected from the proximity to the QSO.

\subsection{Aluminium}

\begin{figure}
\centerline{\rotatebox{0}{\resizebox{9cm}{!}
{\includegraphics{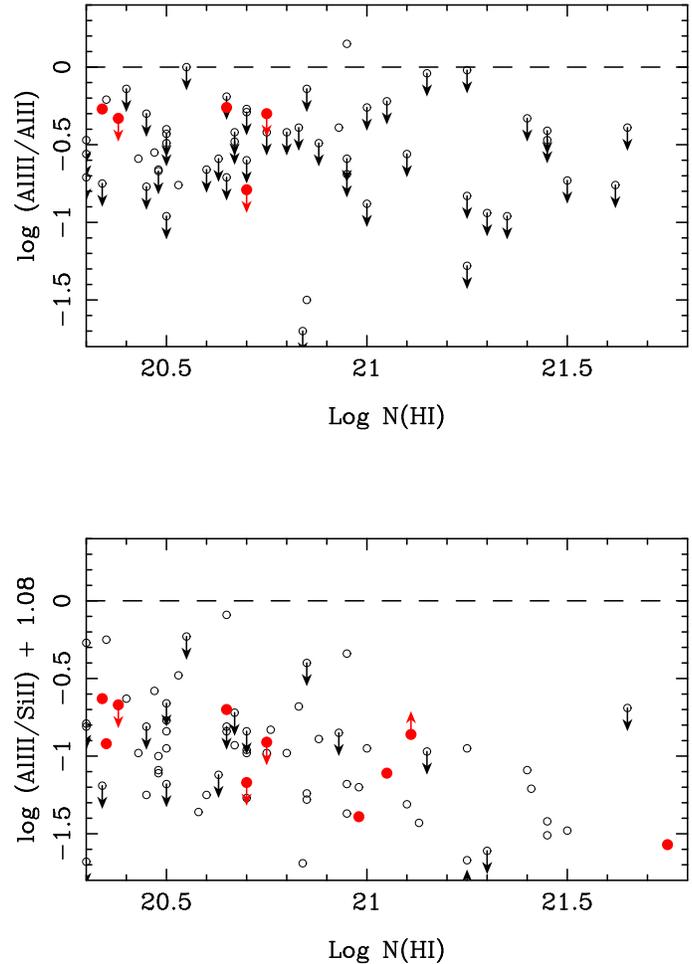}}}}
\caption{\label{Al23} AlIII/AlII (top) and AlIII/SiII (bottom) as a
function of \nhi\ for intervening DLAs (open points, $\Delta V
>10000$ \kms) and PDLAs (filled points, $\Delta V <3000$ \kms).}
\end{figure}

A natural probe of the level of ionization in a DLA is to compare
column densities of different ionization states of a given element.
The relative column densities depend on the ionization potentials, the
shape of the ionizing spectrum and the density of the gas (or
equivalently, the ionization parameter).  The ratio of AlIII/AlII is
sometimes used as a crude estimate of ionization, although SiII is often
substituted for AlII which saturates quickly.  Vladilo et al. (2001)
showed a broad anti-correlation of AlIII/AlII with \nhi\ supporting
the idea that this ratio traces the amount of ionized gas.  In Figure
\ref{Al23} we show AlIII/AlII and AlIII/SiII for our PDLA and
comparison DLA sample, where the latter ratio is corrected for the
solar value of (Si/Al).  There is no obvious difference between the
DLA and PDLA ratios, certainly there is no indication of elevated
amounts of AlIII at a given \nhi\ in PDLAs relative to DLAs.  In fact,
several PDLAs appear to have fairly low AlIII/SiII.  However, as
discussed by Howk \& Sembach (1999) AlIII may actually be a fairly
poor tracer of ionized gas in intervening absorbers and may even 
originate from a different
region in the absorber (Vladilo et al. 2001).  Moreover, the behaviour
of AlIII/AlII is very sensitive to the shape of the ionizing spectrum.
Vladilo et al. (2001) use \textsc{Cloudy} photoionization models to demonstrate
that for a hard (QSO) ionizing spectrum the observed AlIII/AlII is 
essentially flat as a function of \nhi, for a given ionization parameter
(log U). Therefore, whilst adopting a stellar ionizing spectrum seems to 
naturally reproduce the trend of decreasing AlIII/AlII with increasing \nhi\ 
for a given ionization parameter, assuming a hard spectrum requires the
log U to decrease at higher \nhi\ to produce the same anti-correlation
(see also Vladilo et al. 2003).
It requires a very intense power law spectrum (log U $\gtrsim -1$)
in order to achieve N(AlIII) $>$ N(AlII), so the relatively low ratios 
of AlIII/SiII observed in Figure \ref{Al23} do not rule out proximity
to a hard ionizing source.  In fact, a hard ionizing spectrum
may actually be the reason for the lower AlIII/SiII ratios at log 
\nhi\ $<$21 in the PDLAs compared to intervening DLAs.  The stellar
ionizing models of Vladilo et al. (2001) generally predict higher 
AlIII/AlII at low \nhi\ for soft spectra relative to QSO-like radiation.

\subsection{Silicon}

Next, we consider the relative abundances of SiII and SiIV.  SiII is
detected in all 7 PDLAs in our sample.  There are 4 SiIV detections,
two lower limits (from saturated lines) and one upper limit from a
non-detection.  For the two PDLAs with log \nhi\ $<$ 20.4 (Q0151+048
and J0142+0023) we find that log[N(SiIV)/N(SiII)]= $-0.26$ and $-0.42$
respectively.  Such high fractions of SiIV are relatively rare in
intervening DLAs.  The other PDLAs where SiIV has a
detection or upper limit have SiIV/SiII fractions less than 10\%.  In
cases where intervening DLAs exhibit SiIV, it is usually observed in a
different velocity structure from SiII (Wolfe \& Prochaska 2000),
which can be obtained if the ionizing spectrum is relatively soft
(Howk \& Sembach 1999).  This is indeed the case for the PDLAs towards
J2321+1421 and J0142+0023 where the bulk of the SiIV is offset from
the SiII.  However, for 3/6 of the PDLAs in our sample where we detect
SiIV, it traces the structure of the SiII (Q0151+048, J1131+6044,
J1604+3951).  This is unusual; one explanation could be the presence
of a hard spectrum (and/or high ionzation parameter).  For the final
PDLA where SiIV is detected (J1240+1455) the SiIV is so strongly
saturated we have no information on its velocity structure.
Interestingly,  the coincident  (or not) velocity structure of SiIV
with SiII is independent of the $\Delta V$ of the PDLA.  For example,
J2321+1421 has the smallest velocity separation ($-1616$ \kms) with
little or no SiIV coinciding with SiII, yet the largest $\Delta V$
PDLA (J1131+6044, 2424 \kms) has very similar velocity structure in
the two ions.

\subsection{Alpha elements}\label{ion_alpha}

S, Si, Ar and O are all alpha capture elements that are produced
predominantly in massive stars.  Their shared nucleosynthetic origin
leads to approximately solar abundance ratios of these elements in
Galactic stars (e.g. Chen et al. 2002).  Departures from solar ratios in
DLAs can occur when one or both of the alpha elements in question is
depleted from the gas phase onto dust, or if there is significant
ionization.  OI is an ideal `anchor' to study alpha elements since
oxygen is both relatively undepleted (Savage \& Sembach 1991) and OI
requires a negligible ionization correction.  Unfortunately, the most
accessible OI transition ($\lambda_0$ = 1302 \AA) is usually strongly
saturated.  We have only one PDLA in our sample with a
well-constrained column density of OI; the other absorbers yield only
lower limits.  

\begin{figure}
\centerline{\rotatebox{270}{\resizebox{6cm}{!}
{\includegraphics{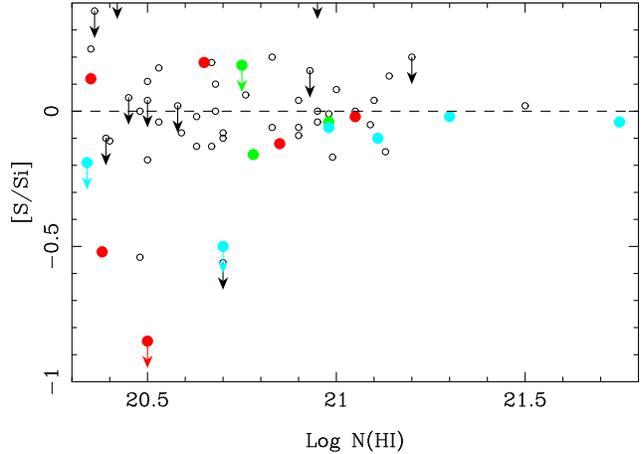}}}}
\caption{\label{S_Si_HI} [S/Si] as a function of
log \nhi\ for intervening DLAs (open points, $\Delta V
>10000$ \kms) and PDLAs (filled coloured points) from the new data
presented here and taken from the literature (see Table 
\ref{lit_abund_table}).  PDLAs are colour
coded by their relative velocities (see Table \ref{dv_table}): cyan for
$\Delta V < 500$ \kms, green for $500 < \Delta V < 1500$ \kms\ and
red for $\Delta V > 1500$ \kms.  The low values of [S/Si] can not be
easily explained by dust depletion or nucleosynthesis and may be
caused by ionization by a hard radiation source.}
\end{figure}

We have detections or meanginful limits for S and Si for all the PDLAs
in our sample.  Like O, S is largely undepleted in the Galactic ISM,
but Si is mildly refractory.  Super-solar ratios of [S/Si] may
therefore be observed if dust is present.  In Figure \ref{S_Si_HI} we
plot [S/Si] versus \nhi\ for the sample of 7 new PDLAs plus systems
taken from the literature (see Table \ref{lit_abund_table}).  PDLAs
are colour-coded by their relative velocity and intervening DLAs with
$\Delta V >$ 10,000 \kms\ are shown for comparison as open circles.
The intervening DLAs cluster around [S/Si]=0 with most points within
$\pm$0.2 dex of the solar ratio.  The effects of dust depletion appear
to be relatively mild (i.e. few DLAs show highly super-solar [S/Si]).
The PDLAs also cluster around the solar ratio, but only when log \nhi\
$\ge$ 20.75.  At least 50\% (accounting for limits) of the lower \nhi\
PDLAs have [S/Si] that is several tenths of a dex below that of the
intervening systems.  Since neither nucleosynthetic nor depletion are
expected to lower the [S/Si], the deviation from the solar ratio may
be due to ionization effects.  This hypothesis is supported by the
tendency of low [S/Si] ratios to be observed only at lower \nhi\
PDLAs.  Photoionization modelling indeed indicates that S and Si
require corrections of opposite sign in the presence of a power law
spectrum (Rix et al. 2007).  Interestingly, the trend of low [S/Si] at
low \nhi\ does not seem
to depend solely on $\Delta V$.  Even at $\Delta V >$ 1500 \kms, 2/3
PDLAs with log \nhi\ $\le$ 20.5 show very sub-solar [S/Si] ratios.
Conversely, PDLAs with very small velocity separations exhibit solar
[S/Si] when log \nhi\ $>$ 21, indicating that they are well-shielded
despite a small velocity separation.  However, we caution that many of
the relative velocities are constrained only to within $\pm$ 500 \kms\
and that $\Delta V$ is not necessarily an indication of physical
proximity.

A natural concern in the interpretation of Figure \ref{S_Si_HI} is
that it relies on several limits in PDLAs resulting from the
non-detection of SII.  As discussed in Section \ref{coldens_sec} the
calculated limit is somewhat sensitive to the assumed width of the
line.  In order to check our calculated SII detection limits we
produced Voigt profiles of SII $\lambda$ 1259 for the limiting column
density, adopting the $b$-value and redshift of the strongest
component in the fit to one of the low ions (either SiII or FeII).  A
visual inspection of the calculated limits overlaid on the data
indicate that they are reasonable estimates.

Argon is relatively rarely measured in DLAs, mostly due frequent
blending of the rest-frame far UV lines of ArI at $\lambda = 1048, 1066$
\AA.  However, the abundance of ArI is very sensitive to ionization,
because its ratio of photoionization to recombination rates are 
typically one order of magnitude higher than for HI (Sofia
\& Jenkins 1998).  Interestingly for our study, Vladilo et al.
(2003) have used \textsc{Cloudy} modelling to demonstrate that  the fraction
of ArI is sensitive to the adopted radiation field.  Low ratios
of Ar relative to other elements are expected when the radiation
field is hard, whereas the models predict solar ratios when the
ionizing spectrum is soft.  Two of our limits on [Ar/Si] are not
deep enough to be very meangingful: [Ar/Si] $<+0.42, -0.01$ towards
J0140$-$0839 and J2321+1421 respectively.  The two other limits are
very sub-solar: [Ar/Si] $<-0.86, -0.47$ towards
J1131$-$6044 and J0142+0023 respectively.  We have one new argon detection
(towards J1604+3951),
although we can only derive its ratio with silicon for the principal component:
[Ar/Si]=$-0.40$.  Sub-solar ratios of [Ar/Si] are therefore apparently
common in the PDLAs and further support significant ionization by a hard
radiation field.

\subsection{Nitrogen}

The most highly ionized species available for study in our sample is
NV (in all cases OVI is blended).  Fox et al. (2009) find a
13$^{+5}_{-4}$\% NV detection rate in DLAs compared with
13$^{+18}_{-9}$\% in PDLAs out to 5000 \kms\ from the QSO.  This
result may seem surprising given the ionization radiation from the
nearby QSO.  Indeed, Fechner \& Richter (2009) find a high NV
incidence rate amongst $\Delta V < 5000$ \kms\ absorbers with 13 $<$
log \nhi\ $<$ 17.0.  Fox et al. (2009) suggest that the consistency in
NV detection rates between DLAs and PDLAs may be due to the fairly
large velocity interval over which they include a DLA in their
proximate sample.  In our sample, 2/7 (J1240+1455 and J1604+3951) of
the PDLAs exhibit NV at a similar velocity to the low ions -- tentative
evidence that NV is more common in PDLAs, although better statistics
are required to confirm this.  Notably,
the PDLA towards J1240+1455 exhibits one of the largest N(NV) yet
reported in the literature (Fox et al. 2009).  The two NV detections
in our sample occur in PDLAs with small $\Delta V$, high \nhi\ and
relatively high metallicity ($Z \sim 1/10 Z_{\odot}$).  Rix et
al. (2007) also detected NV in the \nhi=20.98 PDLA towards
Q2343$-$BX415, which also has a fairly low velocity separation
($\Delta V = 22$ \kms) and high metallicity ($Z \sim 1/5 Z_{\odot}$).
However, J2321+1421 which has the largest negative velocity in our
sample ($\sim -1600$ \kms) does not exhibit NV.  The detection rate of
NV would therefore increase if we limited our statistics to lower
$\Delta V$, supporting the explanation of Fox et al. for the lack of
NV excess in their proximate sample.  The higher incidence of NV out
to 5000 \kms\ seen by Fechner \& Richter (2009) may be associated with
the lower \nhi\ column densities of their sample.

\section{Abundances}\label{abund_sec}

\subsection{Metallicity}\label{metal_sec}

\begin{figure}
\centerline{\rotatebox{0}{\resizebox{9cm}{!}
{\includegraphics{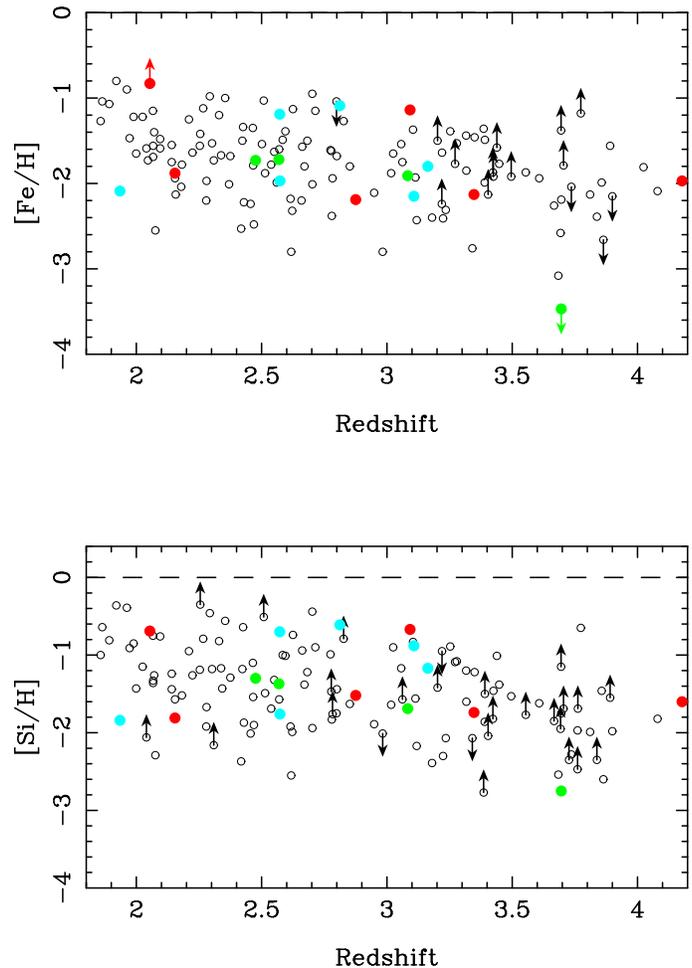}}}}
\caption{\label{Fe_Si_z} Si and Fe abundances as a function of
absorption redshift for intervening DLAs (open points, $\Delta V
>10000$ \kms) and PDLAs (filled coloured points).  PDLAs are colour
coded by their relative velocities (see Table \ref{lit_abund_table}):
cyan for $\Delta V < 500$ \kms, green for $500 < \Delta V < 1500$
\kms\ and red for $\Delta V > 1500$ \kms.  There is a large scatter in
both Si and Fe abundances in PDLAs at a given redshift (see Section
\ref{metal_sec}).}
\end{figure}

\begin{figure*}
\centerline{\rotatebox{270}{\resizebox{12cm}{!}
{\includegraphics{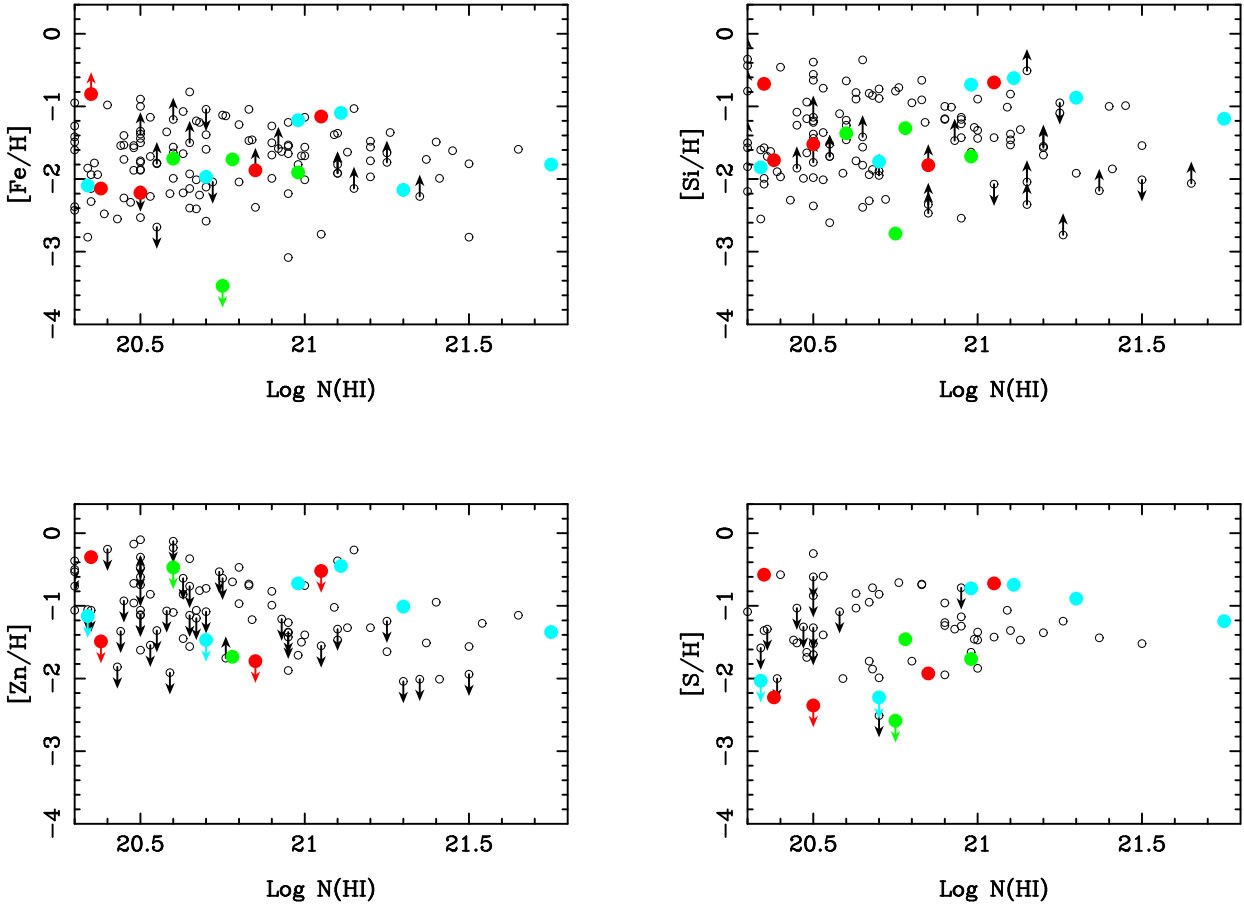}}}}
\caption{\label{Fe_Si_S_Zn_HI} Si, Fe, Zn and S abundances as a function of
log \nhi\ for intervening DLAs (open points, $\Delta V
>10000$ \kms) and PDLAs (filled coloured points). Si and Fe
are either mildly or severely depleted by dust, whereas Zn and S
are largely undepleted from the gas phase and may therefore represent
more reliable indicators of metallicity.  PDLAs are colour
coded by their relative velocities (see Table \ref{lit_abund_table}): cyan for
$\Delta V < 500$ \kms, green for $500 < \Delta V < 1500$  \kms\ and
red for $\Delta V > 1500$ \kms. Silicon and sulphur (and to a lesser
extent, zinc) exhibit high abundances when log \nhi\ $>$ 21 (see Section
\ref{metal_sec}).}
\end{figure*}

Figure \ref{Fe_Si_z} shows Fe and Si abundances as a function of
redshift for DLAs and PDLAs whose redshifts are $z_{\rm abs} > 1.8$.
Our sample includes PDLAs that are amongst both the most metal-poor
and the most metal-rich for their redshift.  The PDLA towards
J0140$-$0839 has the lowest metallicity ever detected in a high
\nhi\ absorber (see Section \ref{0140_sec} for further discussion),
although there may be an upward correction of up to a few tenths of
a dex to be made for ionization.
The results in Figure \ref{Fe_Si_z} ostensibly support the conclusion
of Rix et al. (2007) that PDLAs exhibit metallicities that span the
distribution of intervening DLAs.  However, in Section \ref{ion_sec}
it is argued that the low \nhi\ systems are likely to have significant
ionization corrections.  We therefore re-assess the PDLA metallicities
as a function of \nhi.

PDLAs with log \nhi\ $>$ 21 probably have negligible ionization
corrections and Figure \ref{Fe_Si_S_Zn_HI} shows that these PDLAs have
quite high [Si, S/H].  The Zn abundances are also distributed towards
the upper end of the intervening DLA distribution.
[Fe/H] shows more scatter and its
sensitivity to dust depletion makes it harder to interpret.  At lower
\nhi, all but one of the PDLAs have Si and S abundances that are below
the median.  This intriguing observation could be interpreted as a
difference in the provenance of the low and high \nhi\ PDLAs.
Alternatively, (and perhaps more likely) it could indicate that the
PDLAs (or at least a subset thereof) have intrinsically higher
metallicities, but whose abundance determinations are 
affected by ionization at low \nhi.  

\subsection{Dust depletion indicators}\label{dust_sec}

\begin{figure}
\centerline{\rotatebox{270}{\resizebox{6cm}{!}
{\includegraphics{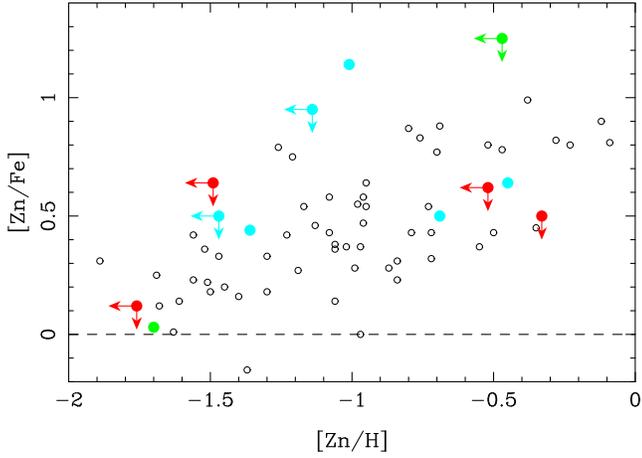}}}}
\caption{\label{Zn_Fe} [Zn/Fe] as a function of [Zn/H] for intervening 
DLAs (open points, $\Delta V >10000$ \kms, limits excluded for clarity) 
and PDLAs (filled coloured points).  PDLAs are colour
coded by their relative velocities (see Table \ref{lit_abund_table}): cyan for
$\Delta V < 500$ \kms, green for $500 < \Delta V < 1500$  \kms\ and
red for $\Delta V > 1500$ \kms.  The PDLAs show no distinct dust depletion
trends compared with the intervening absorbers (see Section
\ref{dust_sec}).}
\end{figure}

In Figure \ref{Zn_Fe} we plot the [Zn/Fe] ratio (as an indicator of
depletion) as a function of [Zn/H] for the PDLA and DLA samples.  The
median [Zn/Fe] for our literature sample (when both Zn and Fe are
measured) is [Zn/Fe] $\sim$ +0.4.  J1240+1455 appears to have
particularly high depletion, although at $Z \sim 1/10 Z_{\odot}$ it is
also one of the more metal-rich absorbers.  The PDLA towards
J1240+1455 has a high \nhi\ so the high [Zn/Fe] ratio is unlikely to
be caused by ionization.  J1604+3951 shows clear signs of different
levels of depletion between two velocity components with high [Zn/Fe]
in one component but not the other (see Section \ref{1604_sec}).
Overall, there is no obvious difference between the dust depletion in
the PDLAs and the intervening systems.

\subsection{Alpha elements}\label{alpha_sec}

\begin{figure}
\centerline{\rotatebox{0}{\resizebox{9cm}{!}
{\includegraphics{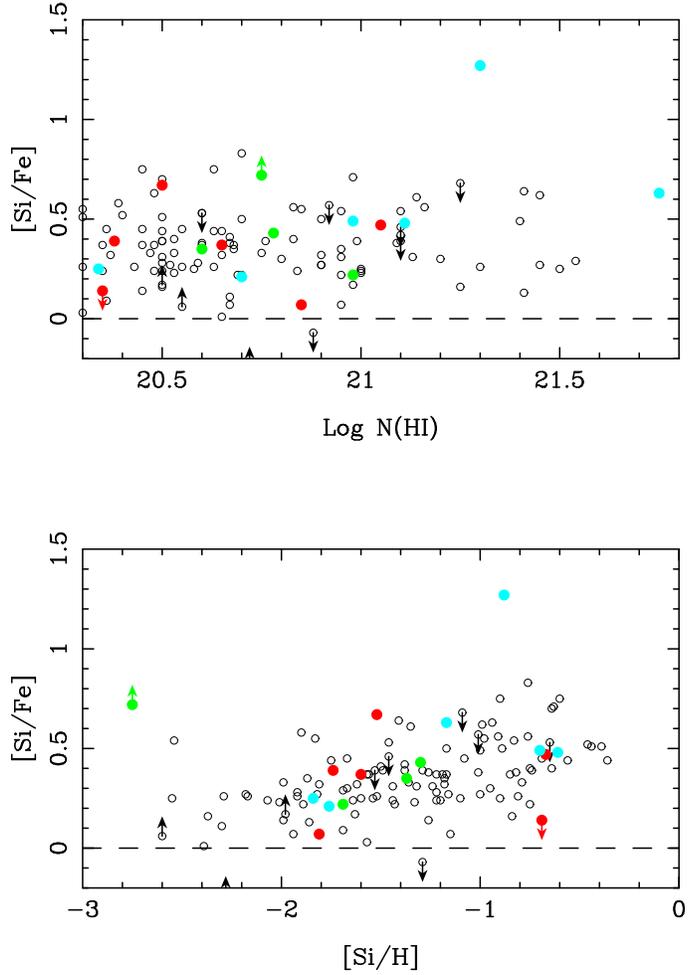}}}}
\caption{\label{Si_Fe_HI} [Si/Fe] as a function of [Si/H] (bottom panel)
and log \nhi\ (top panel) for intervening DLAs (open points, $\Delta V
>10000$\kms) and PDLAs (filled coloured points).  PDLAs are colour
coded by their relative velocities (see Table \ref{lit_abund_table}): cyan for
$\Delta V < 500$ \kms, green for $500 < \Delta V < 1500$  \kms\ and
red for $\Delta V > 1500$ \kms.  The PDLAs show no distinct [$\alpha$/Fe]
trends compared with the intervening absorbers (see Section
\ref{alpha_sec}).}
\end{figure}

[Si/Fe] is used as an indicator of $\alpha$ enhancement, since both
elements are measured in the majority of our sample.  [Si/Fe] is
usually plotted as a fuction of metallicity, although the results
presented so far in this paper indicate that trends may also be
present with \nhi\ due to ionization.  In Figure \ref{Si_Fe_HI} we
therefore plot [Si/Fe] as a function of both \nhi\ and [Si/H].  Figure
\ref{Si_Fe_HI} shows that at low \nhi\ the PDLAs have fairly typical
values of [Si/Fe], with the exception of the very metal-poor PDLA
towards J0140$-$0839 which has [Si/Fe] $\ge$ 0.72.  This PDLA is even
more of an outlier when its [Si/Fe] is considered relative to its
metallicity.  Although the high [Si/Fe] ratio could be due in part to
ionization corrections and/or dust, it also raises the intriguing
possibility of extreme alpha element enhancement in a chemically young
object.  SDSS offers the opportunity to search out other rare, low
metallicity DLAs in order to investigate their nucleosynthetic
histories.  The [O/Fe] ratio in this absorber is also high,
[O/Fe]$\ge$ 0.75 and OI is much less affected by ionization and dust
than SiII.  J1131+6044 also has a relatively high value of
[Si/Fe]=0.67, although Figure \ref{Si_Fe_HI} shows that such values
are not unknown in the intervening population.  J1240+1455 has an
extremely high [Si/Fe] that is unlikely to be due to ionization
effects, given its high \nhi.  However, as discussed in the previous
subsection, the [Zn/Fe] indicates a large depletion fraction which
could lead to an over-estimate of [Si/Fe] (Fe is much more refractory
than Si).  Using the undepleted ratio of [S/Zn]=0.11, the high [Si/Fe]
of J1240+1455 appears to be largely due to dust.  In summary, there
are no clear, systematic differences in the [Si/Fe] ratios of PDLAs
compared with the intervening systems, although some PDLAs do appear
to have relatively high values.

\section{Constraining the distances to PDLAs from fine structure lines}\label{dist_sec}

We have emphasized in this paper that the velocity offsets ($\Delta V$)
between the QSO and PDLAs are unlikely to be convertible into
distances using Hubble's law.  To attempt to constrain the true
physical distances we can model the interplay between the QSO's radiation
and the PDLA's interstellar gas.

\subsection{Distances from CII$^{\star}$}\label{c2_sec}

In this subsection we describe the method for using  C II$^{\star}$
to estimate the PDLA--QSO distance, as well as describing the
assumptions of the model, limitations and caveats.  The practical
application of the method is given in Section \ref{app_sec}. 

Wolfe, Prochaska \& Gawiser (2003) have used CII$^{\star}$ to estimate
the radiation intensity, and hence the star formation rate in DLAs.
The far UV radiation emitted by massive stars contributes to heating
through the grain photo-electric effect, which in turn heats the gas.
The heating rate, $\Gamma$ is a function of the mean radiation
intensity (J), the heating efficiency ($\epsilon$) and the dust to gas
ratio relative to the Galactic value ($\kappa$): $\Gamma \propto
J\epsilon \kappa$.  The value of $\epsilon$ is known from Galactic
studies (e.g. Bakes \& Tielens 1994), and $\kappa$ can be inferred
from ratios of refractory to undepleted elements for individual DLAs,
under an assumed model for dust depletion patterns and intrinsic
abundance ratios.  In a plane parallel layer J is proportional to
$\psi_{\star}$, the star formation rate per unit area (which we henceforth
refer to as simply `star formation rate' for brevity).  The star
formation rate can therefore be determined once $\Gamma$ is known.
This is achieved by assuming that the medium is in local thermal
equilibrium and that heating rate can thus be equated to the cooling
rate.  This latter quantity is inferred from the C II$^{\star}$
absorption line which measures the population of the excited
$^2P_{3/2}$ fine-structure state that spontaneously decays to the
ground-state by emitting a $158 \mu$m photon.

For PDLAs, a second source of far UV photons potentially contributes to
the heating: radiation directly from the proximate QSO.  Rix et
al. (2007) measured CII$^{\star}$ in one PDLA and by assuming that all
the radiation originated from the QSO placed a lower limit on the
distance of the PDLA from the QSO.  We follow a similar procedure
here, but calculate the distance as a function of $\psi_{\star}$ to
explore the range of likely values of the PDLA--QSO separation.  We
assume that the mean radiation intensity, J, inferred from
CII$^{\star}$ has contributions from star formation and the QSO:

\begin{equation}\label{J_eqn}
J_{\rm TOT} = J_{\rm SF} + J_{\rm QSO}.
\end{equation}

\noindent The radiation contribution from the QSO depends on the QSO
luminosity ($L_{\nu}$) and separation (d).  J is measured in ergs
cm$^{-2}$ s$^{-1}$ sr$^{-1}$ Hz$^{-1}$ so that

\begin{equation}\label{J_qso}
4\pi J_{\rm QSO, \nu} = \frac{L_{\nu}}{4\pi d^2}.
\end{equation}

\noindent  The CII$^{\star}$ analysis of Wolfe et al. (2003) yields 
J for a rest wavelength of 1500 \AA.  

Combining equations \ref{J_eqn} and \ref{J_qso} and solving for $d$,
we obtain

\begin{equation}\label{d_eqn}
d = \frac{1}{4\pi}\sqrt{\frac{L_{\nu}}{J_{\rm TOT, \nu}-J_{\rm SF, \nu}}}
\end{equation}

\noindent The contribution to J from star formation depends on the
geometry of the plane parallel layer (parameterised by the ratio of
its radius and height, R/h), metallicity and its dust-to-gas ratio,
$k_{\nu}$.  For $k_{\nu}h \ll k_{\nu}R \ll 1$ equations 18 and 19 of
Wolfe et al. (2003) can be reduced and re-arranged to give

\begin{equation}\label{J_sf}
J_{\rm SF, \nu} = 8.4 \times 10^{-16} \psi_{\star} . \frac{1+ln(R/h)}{8\pi}
\end{equation}

\noindent  where $\psi_{\star}$ is measured in units of M$_{\sun}$ yr$^{-1}$
kpc$^{-2}$.  A fixed aspect ratio of R/h=20 is used for all
calculations.  Following Wolfe et al. (2003), the dust-to-gas ratio is
determined from Si/Fe and metallicity from [Si/H] with an assumed
intrinsic [Si/Fe]=0.2 and SMC depletion patterns.  Solutions can be
calculated for two cases: gas that is dominated by the cold
and warm neutral media (CNM, WNM) respectively.   However, using
CII$^{\star}$ to constrain distances for the WNM solutions has two
problems.  First, in the WNM, the cooling is actually dominated not by
the [CII] 158 $\mu$m transition, but by \lya\ and the recombination of
electrons onto grains.  Second, at the low densities implied by the
WNM solutions, the dominant heating source is no longer the grain
photo-electric effect, but cosmic ray heating and, to a lesser extent,
heating by X-rays.  In the calculations of Wolfe et al. (2003), the
cosmic ray production rate is assumed to scale with the star formation
rate.  However, we are additionally concerned with energy sources
associated with the QSO and it is unclear whether AGN produce cosmic
rays and at what rate.  We therefore only consider the CNM solutions,
but caution that these are probably inappropriate for the PDLAs with
the lowest cooling rates.  A full treatment of the WNM solutions for
AGN heating is beyond the scope of this paper.

The QSO luminosity at $\lambda_0 = 1500$ \AA\ is determined from the
SDSS spectrum, except in the case of Q0151+048.  For this target,
we take B$_{AB}$=17.83 from Fynbo et al. (2000) which converts to a flux of
2.7$\times 10^{-27}$ ergs s$^{-1}$ Hz$^{-1}$ (no corrections
are made for \lya\ emission or \lya\ forest absorption, since
they lie outside the $B$ filter for $z=1.9$).
The luminosity and flux (in units of ergs s$^{-1}$ Hz$^{-1}$) at a
given frequency are related through the equation

\begin{equation}
F_{\nu} = L_{(1+z)\nu}\times \frac{1+z}{4\pi d_L^2}
\end{equation}

\noindent  where $d_L$ is the luminosity distance in our adopted cosmology.
In Table \ref{lum_table} we list the calculated values of $L_{1500}$.

In previous sections, we have argued that ionization may lead to
errors in in the column densities that we measure, particularly at low
\nhi.  The metallicity (nominally [Si/H]) and dust to gas ratio
(calculated from [Si/Fe]) are required to calculate $J_{TOT}$.
However, Wolfe et al. (2003) have shown that $J_{TOT}$ is actually
relatively insensitive to the metallicity used for values $Z \lesssim
1/10 Z_{\odot}$.  Even at the highest metallicities, incorrect values
of [Si/H] are likely to affect $J_{TOT}$ by less than 20\%.
Uncertainties in the dust-to-gas ratio are potentially more
problematic, since the grain photo-electric effect can be a dominant
source of ISM heating.  However, the contribution of this process to
the heating budget depends on the ISM density (e.g. Figure 3 in Wolfe
et al. 2003), becoming more dominant at high densities.  For CNM
solutions, lower dust-to-gas ratios (i.e. fewer grain targets for
heating) requires a higher incident intensity to account for the
observed CII$^{\star}$\footnote{As discussed in the text, at low
densities cosmic ray heating dominates over the grain photo-electric
effect.  WNM solutions, although not pursued here, are therefore very
insensitive to the input value of [Si/Fe].}.  However, in Section
\ref{alpha_sec} we showed that the range of [Si/Fe] ratios of most of
the PDLAs are in good agreement with intervening systems.  Assuming
that there is an intrinsic `floor' to [Si/Fe] whose value is
approximately 0.2 (e.g. Prochaska \& Wolfe 2002) and given that
ionization effects tend to lead to over-estimates of [Si/Fe]
(e.g. Dessauges-Zavadsky et al. 2003; Howk \& Sembach 1999), the
measured Si/Fe ratios do not indicate significant corrections in the
majority of cases.

\subsubsection{Application to the data}\label{app_sec}

\begin{figure}
\centerline{\rotatebox{270}{\resizebox{6cm}{!}
{\includegraphics{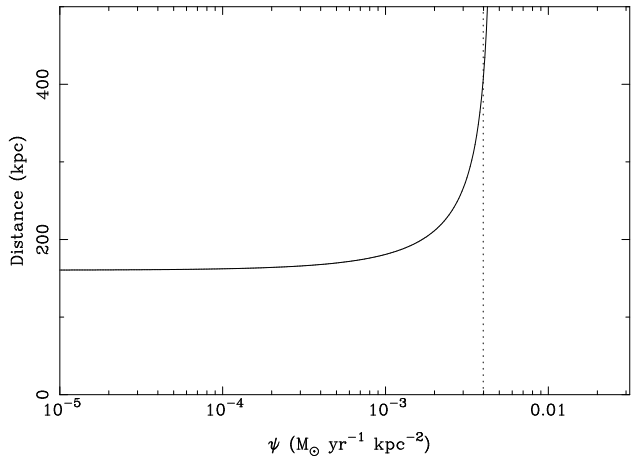}}}}
\caption{\label{sfr_dist} Distances between the PDLA and Q0151+048
calculated from the CII$^{\star}$ analysis.  The curve shows the distance
corresponding to an increasing local contribution of radiation from
star formation rate, log $\psi_{\star}$.  The vertical dotted line
indicates log $\psi_{\star} = -2.4$ M$_{\odot}$ yr$^{-1}$.
kpc $^{-2}$.  }
\end{figure}

\begin{center}
\begin{table*}
\begin{tabular}{lcccccccc}
\hline
QSO & $z_{\rm abs}$ & $\Delta V$ & log N(CII$^{\star}$) & log N(SiII$^{\star}$) & L$_{1500}$    &  d$_{\rm CNM}$ (kpc) & d$_{\rm CNM}$ (kpc) & d$_{\rm SiII^{\star}}$ \\
  & & (\kms)  & (\cm)  & (\cm)  & (ergs s$^{-1}$ Hz$^{-1}$)   & $\psi_{\star}$ = 0 & log $\psi_{\star} = -2.4$ &  (kpc) \\
\hline
J0140$-$0839 &  3.6960 & 1250  &$<$12.41 & $<$11.48&  1.08$\times 10^{32}$ & $>$2099.0 & ... &$>$14.8 \\
J0142+0023 & 3.34765 & 1772 & ... & $<$11.17 & 4.76$\times 10^{31}$ & ... & ... & $>$29.8 \\ 
Q0151+048 & 1.9342 & $-1199$ & 13.0 & $<$ 11.81 & 2.44$\times 10^{31}$ &160.5 &406.0 & $>$8.6 \\
J1131+6044 & 2.8754 & 2424 & $<$12.51 & $<$11.57 & 5.85$\times 10^{31}$ &... & ... & $>$30.8\\
Q2321+1421 & 2.5731 & $-1616$& $<$12.55 &$<$11.78 &  2.27$\times 10^{31}$& $>$510.1 &...& $>$14.3\\
\hline 
\end{tabular}
\caption{\label{lum_table}PDLA--QSO distance calculations.  Solutions are given
for the CII$^{\star}$ analysis under the assumption of a CNM and for SiII$^{\star}$
where suitable detections/limits are possible (see text for details).}
\end{table*}
\end{center}

CII$^{\star}$ is detected in Q0151+048 and J1604+3951 and we have
upper limits for J0140$-$0839, J1131+6044 and J2321+1421 which will
yield lower distance limits for a given $\psi_{\star}$.  However, the
CII$^{\star}$ towards J1604+3951 is partly blended with CII $\lambda$
1334, and the unblended components are saturated, so we do not
consider this PDLA further.  Q0151+048 may have $J_{\rm QSO}$
contributions from Q0151+048B, introducing a second unknown distance.
Fynbo et al. (1999) have argued that the most likely orientation of
the system is that Q0151+048B is in front of Q0151+048 and the DLA and
causes \lya\ emission on the near face of the absorber.  Modelling
this system completely is complicated not just by uncertainty in the
geometry of the QSO pair and absorber, but also by the anisotropy of
quasar radiation and hence the flux `seen' by the DLA coming from
Q0151+048B.  However, assuming that all the flux comes from the
background QSO still yields a lower limit to the distance between
Q0151+048 and the PDLA (just as we determine a lower limit under the
assumption that internal star formation does not contribute).  

In Table \ref{lum_table} we list the limiting value of the PDLA--QSO
distance for a star formation rate $\psi_{\star} = 0$ and also the
fiducial case of a Galactic star formation rate of log $\psi_{\star} =
-2.4$ M$_{\odot}$ yr$^{-1}$ kpc $^{-2}$ for the CNM solutions.  For
PDLAs with upper limits to N(CII$^{\star}$) the CNM solutions are
likely to be inappropriate and they are given in the Table only for
completeness.  More sensitive CII$^{\star}$ limits could definitively
rule out CNM solutions entirely as is the case for J1131+6044 which
only has a WNM solution.  The CII$^{\star}$ analysis therefore only
yields useful a distance constraint for Q0151+048 where the lower
limit is 160 kpc (no star formation case), rising to just over 400 kpc
for a Galactic star formation rate, a separation approximately 5 times
larger than the only other PDLA in the literature has been analysed in
this way (Q2343$-$BX415 by Rix et al. 2007).  In Figure \ref{sfr_dist}
we show the complete range PDLA--QSO distances as a function of
star formation rate for Q0151+048.  The Figure demonstrates that once
star formation dominates over the QSO's radiation, the inferred
distance to the PDLA rises dramatically, essentially providing an
upper limit to the likely star formation rate in the galaxy, as well
limits of the distance to the QSO.

\subsection{Distance constraints from SiII$^{\star}$ limits}\label{si_sec}

\begin{figure}
\centerline{\rotatebox{270}{\resizebox{6cm}{!}
{\includegraphics{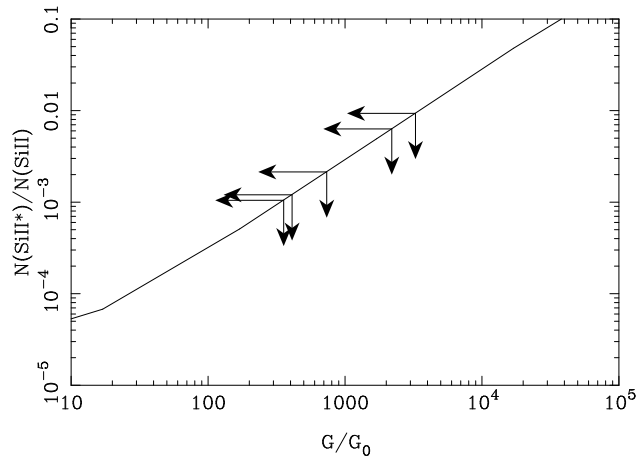}}}}
\caption{\label{pop_fig} Ratio of fine structure to ground state
column densities from UV pumping determined from the \textsc{popratio}
software.  The arrows indicate the inferred upper limit of the
incident radiation field in units of the Galactic value based on upper
limits of N(SiII$^{\star}$).}
\end{figure}

There are a number of uncertainties and assumptions that underpin
the distance calculations from CII$^{\star}$ which affect the
inferred radiation field to varying extents.  These include 
knowledge of the gas phase (cold or warm), metallicity, dust depletion,
assumed geometry and extinction law.  An alternative assessment
of the distance between the absorber and the radiation source can
be made using the SiII fine structure lines and the ratio of
N(SiII$^{\star}$) to the column of the ground state ions.  In order
to determine the relation between N(SiII$^{\star}$)/N(SiII) and the
incident radiation field, we use the publicly available code \textsc{popratio}
(Silva \& Viegas 2002).  We assume that SiII$^{\star}$ is populated purely by
UV-pumping (fluorescence) and ignore contributions from direct (IR)
excitation and collisions.  The results from \textsc{popratio} are shown
by the curve in Figure \ref{pop_fig} as a function of the radiation
field in Galactic units ($G_0$).  Upper limits for N(SiII$^{\star}$)
are derived from SiII$^{\star} \lambda$ 1264 which in turn yield
upper limits on $G/G_0$, which is related to $J$ 

\begin{equation}
J=G \times 1 \times {\rm10^{-19} ~ergs~ cm^{-2} ~s^{-1} ~sr^{-1} ~Hz^{-1}.}
\end{equation}

The luminosities in Table \ref{lum_table} are then combined with $J$
in equation \ref{J_qso} to give a lower limit on the distance.  These
values are lower limits not just because of the non-detection of
SiII$^{\star}$, but also because of our assumption that only
UV-pumping contributes to the population of the fine structure level.
The lower limits derived are typically 15 to 30 kpc, see Table
\ref{lum_table}.  Although these are considerably less stringent than
the CII$^{\star}$ distances, they are fairly robust and depend on very
few assumptions and input parameters.  The only caveat to this
analysis is that at least two of the SiII transitions from the
ground-state must be optically thin (Sarazin, Rybicki \& Flannery
1979), which holds for all but one of our PDLA sample (which is
excluded from this analysis).

\medskip

In summary, we have used two methods to constrain QSO--PDLA distances.
The CII$^{\star}$ method (under the CNM assumption) gives a lower limit
of 160 kpc for Q0151+048 (assuming no internal star formation),
but is rather dependent on model assumptions.  The SiII$^{\star}$ model
is more robust but gives less stringent limits, typically $>$ 15--30 kpc.
Adding star formation to either method increases the inferred distance
of the PDLA from the QSO.  The distances imply that the absorbers
are external to the QSO.

\section{Discussion}\label{discussion_sec}

\subsection{Sulphur and argon as indicators of QSO proximity}

We have suggested that sub-solar ratios
of [S/Si] at low values of \nhi\ may be caused by the significant
under-estimate of N(S) from SII and, to a lesser extent, an
over-estimate of N(Si) from SiII in the presence of a hard ionizing
spectrum. This result could be predicted from the photoionization models
of Rix et al. (2007), who showed that the observed
under-estimate of sulphur is strongly dependent on the ionization
parameter, $U$.  Most models of intervening DLAs and sub-DLAs have
concluded that the value of log $U$ is typically $<-3$ (e.g Howk \&
Sembach 1999; Dessauges-Zavadsky et al.  2003).  However, for absorbers
close to a QSO, the situation can be quite different.  For example,
Prochaska \& Hennawi (2009) find that for a hydrogen volume density of
0.1 atoms cm$^{-3}$ the ionization parameter of one of their transverse
sub-DLAs at a distance of $\sim$ 100 \hkpc\ from the QSO is log $U =
-1.5$.  The QSOs in our sample are typically a factor of 10 brighter
than the case studied by Prochaska \& Hennawi (2009), so ionization
parameters in the range $-2 < \log U < 0$ are quite feasible.

Sub-solar [S/Si] ratios may therefore be used as a
signpost of a nearby hard radiation source.  Two of the intervening
DLAs (FBQS 2334$-$0908 and PSS 0133+0400) in our literature sample
also have [S/Si]$< -0.5$.  One possible explanation could be that
although they are intervening ($\Delta V \gg$ 10,000 \kms) systems,
there is a nearby foreground QSO at a similar redshift.  FBQS
2334$-$0908 is covered by the SDSS footprint.  Although none of the
objects near to FBQS 2334$-$0908 were targeted by the SDSS for
spectroscopy, there are two point sources at separations of 27 and 105
arcseconds (208 and 809 \hkpc\ at $z=3$ respectively) whose colours
are consistent with expectations of QSOs.  PSS 0133+0400 is not
covered by the SDSS, but the APM catalogue shows 2 point sources with
similar colours to the QSO at separations of 57 and 64 arcseconds (409
and 459 \hkpc\ at $z=3.7$ respectively).  It would be interesting to
obtain spectra of these sources to determine whether or not they are
indeed QSOs at the same redshift as the intervening DLAs.  Indeed,
some of the intervening NV absorbers identified by Fox et al. (2009)
have nearby QSOs at the same redshifts (G. Worseck, private
communication).

The low ratios of [Ar/Si] (where measured) in our PDLAs are
also indicative of a hard ionizing spectrum (Vladilo et al. 2003).
Using a sample of 10 ArI measurements, Vladilo et al. (2003) have also
found low ([Ar/Si]$<-0.5$) at $z_{\rm abs}<3$.  At higher redshifts,
the values are close to, or at, the solar level.  Vladilo et
al. (2003) have argued that this may be evidence for evolution in the
ionizing background, which is shifting from a softer to harder shape
as the redshift decreases.  However, 3/7 of the $z_{\rm abs} < 3$
absorbers in the Vladilo et al. (2003) sample are PDLAs.  A further
two have velocities within 10,000 \kms, leaving only two which might
be considered truly intervening absorbers.  At $z_{\rm abs} > 3$ all 3
absorbers have $\Delta V > 15,000$ \kms.  Given the evidence presented
in this paper that the effects of a hard ionizing spectrum are seen
out to at least 3000 \kms\ (and previous results that indicate narrow
associated absorbers may contribute out to 10,000 \kms, e.g. Wild et
al. 2008; Tytler et al. 2009), the apparent redshift evolution may
actually arise from the inclusion of proximate systems.  A larger
sample of large $\Delta V$ DLAs with argon measurements is necessary
to explore more fully the redshift evolution of [Ar/Si].

\subsection{NV as a tracer of high ionization gas}\label{disc_high}

High ionization species such as NV and OVI might be expected to
be more common in absorbers close to the QSO.
Although OVI is blended in almost all of our sightlines, we
have detections and upper limits of NV for all of our PDLAs.
It is intriguing that the two detections of NV
\textit{at the same velocity as the singly ionized metal lines} are
towards the two highest \nhi\ absorbers (towards J1240+1455 and
J1640+3951); both have log \nhi $>$ 21.  However, both absorbers also
have normal [S/Si] ratios, whereas we might expect departures from the
solar ratio if the gas is partly ionized by a hard spectrum (as is
seen for low \nhi\ PDLAs).  The presence of NV in an ISM that is
dominated by neutral gas indicates that the NV might be formed through
an internal, localised source of ionizing radiation.  In that case,
the radiation from the QSO may not be responsible for the formation of
the NV and the proximate nature of the DLA is mis-leading.  Fox et al.
(2009) found that the NV detection rate is higher for higher
metallicity (intervening) DLAs. J1240+1455 and J1640+3951 have
(undepleted) S abundances that are more than 1 dex higher than many of
the other PDLAs in our sample and 3 times higher than intervening DLAs
at the same \nhi\ and redshift.  Fox et al. (2009) have suggested that
active star formation could be responsible for highly ionized gas
which is consistent with the high metallicities, large velocity
spreads and (for J1604+3951) the very high CII$^{\star}$ column
density.  Indeed, Lehner et al.  (2008) found that a pure QSO (hard)
spectrum always under-produced the observed amount of NV in their
study of a DLA with multi-phase gas; adding a soft stellar component
significantly increased the predicted NV column density.  Our own
\textsc{Cloudy} models indicate that even in the presence of a hard radiation
field that is sufficient to produce sub-solar [S/Si] the NV/NI
fraction is still only on the order of 10\%, which is consistent with
the NV column density limits in our lower \nhi\ systems.

Both J1240+1455 and J1640+3951 show additional (and in the former
case, much stronger) NV offset by several hundred \kms\ to the blue.
A third PDLA (towards Q0151+048) shows highly offset NV (by $-825$
\kms) but no NV at zero velocity; in this case, the \nhi\ is much
lower, only log \nhi\ = 20.34.  It is notable that every PDLA in our
sample of 7 echelle spectra (with the exception of J0140$-$0839, which
has very weak metal lines in general) has high ionization gas that is
offset to negative velocities.  In many cases, there is high
ionization gas (such as CIV) at the redshift of the low ions but no
low ions are seen to accompany the negative velocity high ion
components.  In some cases the velocity offset is small (e.g. CIV at
$\sim -100$ \kms\ in J0142+0023).  In other cases, the velocity
offsets exceed $-$500 \kms\ (e.g. Q0151+048).  We summarise these
offset components in Table \ref{high_ion_tab} (see Sections
\ref{0142_sec} to \ref{2321_sec} for more detailed descriptions)
listing both the $\Delta V$ of the PDLA from the background QSO and
the velocity offset of the high ions from the main low ion components.
Rix et al. (2007) also found a highly ionized component (observed in
NV, CIV and SiIV) offset by $\sim$ +500 \kms\ from the PDLA towards
Q2343$-$BX415.  Weidinger et al. (2005) find NV in 2 absorbers
offset from a proximate Lyman limit system by a few thousand
\kms.  In these more extreme cases (e.g. Q0151+048) the absorption
is likely to be unconnected with the proximate HI system and more
akin to the associated high ionization associated systems studied
by D'Odorico et al. (2004) and Fechner \& Richter (2009).

For J1131+6044 and Q0151+048 where the high ion velocity offset is
very high, it is likely that the absorption is arising in gas outside
the PDLA.  As discussed in Section \ref{0151_sec}, Q0151+048 has a
fainter companion separated by 27.5 \hkpc\ and $\sim$ +120 \kms.
Although there is no confirmed companion for J1131+6044 in the NASA
Extragalactic Database (NED), the SDSS image shows a bright ($g =
17.59$) point-source 5.66 arcseconds to the north-west.  The object
has an almost identical $g-r$ colour to J1131+6044 (0.41 and 0.38
respectively), so that this may be a binary QSO.
The putative companion may contribute to the significant ionization
seen in the J1131+6044 PDLA (e.g. high SiIV fraction), despite the
large positive velocity offset from the QSO (2424 \kms, but see below
for further discussion on the interpretation of velocities).  The
other 5 QSOs in our sample have no obvious bright companion within 20
arcseconds on the SDSS image.  If the companion to J1131+6044 is
confirmed, this would mean that 2/7 of the QSOs with PDLAs in our
sample have a close companion.  Q2343$-$BX415 also has a companion,
and its PDLA also has 2 distinct velocity components in the highly
ionized gas, one of which is coincident with the bulk of the low ions,
and the other offset by $\sim$ +500 \kms.

\begin{center}
\begin{table*}
\begin{tabular}{lcccc}
\hline
QSO           & log N(HI) & $\Delta$V$_{\rm HI}$  &  Velocity offset & High\\
 & &  (\kms) &  (\kms) & Ions\\
\hline 
J0142$+$0023  & 20.38$\pm$0.05 &  1772 & $-100$ & CIV, SiIV \\
Q0151+048 & 20.34$\pm$0.02 & $-1199$ & $-825$ & NV, CIV \\
J1131+6044 & 20.50$\pm$0.15 &  2424 &$-800$ & CIV,OVI \\
J1240+1455 & 21.3$\pm$0.2 & 102 & $-125$ & NV \\
J1604+3951 & 21.75$\pm$0.2 & $-656$ & $-400$ & CIV, SiIV, NV \\
J2321+1421 & 20.70$\pm$0.05 & $-1616$ & $-100$ & CIV, SiIV \\
\hline 
\end{tabular}
\caption{\label{high_ion_tab}High ionization species offset from low ion detections.
No offset high ions are detected for J0140$-$0839}
\end{table*}
\end{center}

\subsection{The origin and nature of the PDLAs}

The question that underpins the research of all associated absorbers
is whether they are intrinsic to the QSO (host or outflow) or simply
nearby in velocity space.  M\o ller et al. (1998) examined a number of
hypotheses for the nature of the PDLAs and favour a model in which the
PDLAs are similar in nature to the intervening absorbers, but possibly
located in a preferential environment, such as in the same overdensity
as the QSO.  Rix et al. (2007) suggest that the PDLA towards
Q2343$-$BX415 (included in our literature sample) may be associated
with outflowing material from the quasar host.  Indirect clues to the
provenance of the PDLAs may be garnered from their statistical
properties.  An excess of PDLAs relative to intervening systems has
been confirmed by three studies (Ellison et al. 2002; Russell et
al. 2006; Prochaska et al 2008b).  Although Ellison et al. (2002) found
an additional excess of PDLAs in their radio-selected sample relative
to optically selected QSOs, Russell et al. (2006) found an equal PDLA
enhancement towards radio-loud and radio-quiet QSOs.  The discrepancy
may be due to the limited statistics of the Ellison et al. study, but
a second possibility is the nature of radio sources.  The CORALS
sample used by Ellison et al. (2002) is comprised entirely of (rarer)
flat-spectrum quasars that have compact morphologies.  The sources
detected at 20-cm in the Russell et al. (2006) sample will have a
range of spectral indices and orientations.  It would be interesting
to investigate the dependence of PDLA incidence as a function of radio
spectral index, as has been done extensively for CIV absorbers.  There
is certainly evidence that the quasar's radiation affects the ability
of a PDLA to survive.  Hennawi \& Prochaska (2007) have shown that the
incidence of transverse DLAs in projected QSO pairs over-predicts the
incidence of proximate absorbers by a factor of 4 -- 20.  These
authors suggested that line-of-sight PDLAs are preferentially
photo-evaporated by the QSO's beamed radiation.  A similar
overabundance of transverse MgII absorbers relative to the line of
sight has been seen by Bowen et al. (2006).    Moreover, Prochaska et
al. (2008b) show that despite the higher incidence of PDLAs relative to
the intervening population, a clustering analysis indicates that they
are nonetheless underabundant relative to the expected number density
of galaxies near QSOs.

One of the observations that supports an intrinsic origin for many
associated absorbers, is their solar or super-solar metallicity (see
the Introduction).  We do not find any evidence for such elevated
metallicities in our sample of PDLAs.  The same is true of the
associated MgII systems (some of which may be DLAs) studied by vanden
Berk et al. (2008).  Nonetheless, as discussed above, we do find that
PDLAs with high HI column densities have a mean metallicity that is
higher than intervening systems by around a factor of three.  For
example, [S/H]=$-0.88\pm0.24$ for the proximate absorbers versus
$-1.41\pm0.20$ for those at $\Delta V > 10,000$ \kms.  The correlation
between stellar mass and gas-phase metallicity that exists out to
redshifts of at least 3 (Tremonti et al. 2004; Savaglio et al. 2005;
Erb et al. 2006; Maiolino et al 2008) indicates that the PDLAs might
therefore also be relatively massive. QSOs themselves are highly
biased tracers of mass and are often located in or near galaxy
clusters or other overdensities, although they apparently eschew the
centres of clusters (Sanchez \& Gonzalez-Serrano 1999; Barr et
al. 2003; Sochting, Clowes \& Campusano 2004; Kauffmann et al. 2004;
Lietzen et al. 2009; Hutchings, Scholz and Bianchi 2009).  Since the
typical stellar mass of a galaxy tends to increase with the galaxy
density of its local environment (e.g. Baldry et al. 2006), the
combination of clustering and the mass-metallicity relation may
explain the higher metallicities of PDLAs.  Further evidence for this
scenario comes from the discovery of a likely companion QSO to
J1131+6044, and with the previously known companions of Q0151+048 and
Q2343$-$BX415 (M\o ller et al. 1998; Rix et al. 2007).  Boris et
al. (2007) find evidence that 3/4 of the QSO binaries in their sample
are associated with rich clusters at $z \sim 1$.  Other evidence for
absorbers clustered around QSOs comes from the incidence of transverse
absorbers.  In their study of close projected pairs (in which it is
argued that transverse absorbers suffer less photo-evaporation than
line of sight absorbers) Hennawi et al. (2006) find a 50\% covering
fraction of sub-DLAs within 150 kpc.

It may seem suprising that absorbers with velocity offsets of up to
3000 \kms\ could be considered associated with the QSO environment
when this velocity corresponds to a Hubble flow distance of $\sim$ 10
Mpc (proper) at $z \sim 3$.  However, the velocity offsets are
unlikely to indicate Hubble flow distances, as clearly demonstrated by
the existence of absorbers with very negative velocities.  Although we
have additionally highlighted the uncertainty in the emission
redshifts, these are unlikely to be incorrect by the $>$ 1500 \kms\
required to make all of our absorber offsets positive.  We have
discussed above some of the observations that have shown that QSOs
often occupy overdense regions, a statement that might lead us to
imagine cluster-like structures today with sizes of a few Mpc.
However, Lietzen et al. (2009) have argued that QSOs can also trace
out other overdense structures, such as filaments whose dimensions can be 
tens of Mpc.  Haines et
al. (2004) have found evidence for an extreme structure of galaxies
clustered around quasars on scales of tens of Mpc at $z \sim 1$.

The PDLAs appear to share some characteristics in common with
absorbers associated with gamma-ray burst (GRB) hosts.  The GRB DLAs
have higher metallicities and are skewed to higher \nhi\ than
intervening QSO DLAs (e.g.  Prochaska et al. 2007a).  Higher
[$\alpha$/Fe] are also found in the GRB DLAs, similar to some of the
PDLAs in our sample.  However, the difference between the GRB DLAs and
intervening systems is often attributed to the GRB
sightlines preferentially intersecting the galaxy at smaller
galactocentric radii, rather than a fundamental difference in the
populations probed by the GRBs and QSOs.

Regardless of their origin, if the metallicities
of  (at least the high \nhi) proximate DLAs are relatively high, 
as a population they present
an interesting new selection technique for identifying the most
metal-rich galaxies at high redshift.  With metallicities ranging
from 1/3 to 1/10 of the solar value, the high \nhi\ PDLAs have
metallicities similar to the `metal strong DLAs' (MSDLAs, Herbert-Fort
et al. 2006) at $z \sim 2$ (Kaplan et al., in preparation).
Indeed, a number of the MSDLAs are also PDLAs (Kaplan et al.,
in preparation).  Although this is still more metal-poor than the 
majority of Lyman Break Galaxies (LBG) studied at this redshift, 
most LBG abundances are limited to fairly massive galaxies (Erb et al.
2006).  Gravitational lensing permits studies of fainter galaxies.
Only a handful of such objects have been studied so far, but the
results indicate that galaxies with $\sim L_{\star}$ luminosities
at $z \sim 2$ to 3 have metallicities of $Z \sim 1/2 Z_{\odot}$
(e.g. Teplitz et al. 2000; Pettini et al. 2002; Quider et al. 2009a,b).
If the PDLAs follow a similar mass-metallicity relation, then they
may be only slightly less massive than these lensed LBGs.  High
metallicity systems present a number of interesting possibilities
for studying the high redshift ISM.  For example, Noterdaeme et al.
(2008) have shown that molecular hydrogen is highly dependent on
metallicity.  High metallicities
also present the opportunity to detect and study the abundances
of rarely detected elements such as Ge, B, Cl and Co (Ellison, Ryan
\& Prochaska 2001; Prochaska et al. 2003c).

\subsection{Future Work}

Hennawi et al. (2006) presented a sample of close, projected QSO pairs
where the background quasar exhibits optically thick \lya\ absorption
at the redshift of the foreground quasar.  It is shown that 50\% of
projected QSO pairs have an absorber with log \nhi\ $>$19 at the
redshift of the foreground QSO when the separation is $<$ 150 kpc.
This much higher incidence of transverse absorbers, relative to
proximate line of sight absorbers, is suggested by Hennawi \&
Prochaska (2007) to be due to differential photo-evaporation by
anisotropic radiation of the foreground QSO.  
In the future, it will be interesting to compare the chemical
abundances and ionization indicators of PDLAs to transverse DLAs
at a given \nhi\ (e.g. Prochaska \& Hennawi 2009) and to test fine structure
distance estimates against the measurable transverse separations.

Despite their external provenance, the
QSO's radiation can apparently affect PDLAs out to at least 2500 \kms\
from the QSO.
With a larger sample, and accurate redshifts, it will also be possible
to look for trends with velocity.  Ultimately, this may require the
incorporation of other parameters such as QSO luminosity, including
radiation from nearby companions.  A campaign is currently underway to
obtain IR spectra for a sample of QSOs with PDLAs (including those
presented in this paper) that will yield redshifts accurate to $\sim$
40 \kms\ from the [OIII] $\lambda$ 5007 line.  It is worth noting that
only a few tens of $z>2$ QSOs have redshifts [OIII] $\lambda$ 5007
redshifts (e.g. Scott et al. 2000).  For objects in the Hubble flow,
velocities correspond to distances.  For positive values of $\Delta V$
we can not distinguish between these expansion velocities and peculiar
motions (as expected if the galaxies and QSOs are in the same
gravitational potential.  However, negative $\Delta V$ can not be due
to Hubble expansion, since it would imply the absorber is more distant
than the QSO, which is clearly impossible.  The negative side of the
velocity distribution of PDLAs therefore yields the `true' velocity
distribution of the PDLAs around the QSO, without contamination from
the Hubble flow.  Larger samples will also allow us to study trends
of the combined effect of $\Delta V$ and QSO luminosity.

\section{Conclusions}

We have presented new high resolution echelle spectra for seven
proximate damped Lyman alpha systems with $\Delta V < 3000$ \kms.  The
metal column densities derived from Voigt profile fits and the
apparent optical depth method are complemented with abundances for a
further 9 PDLAs taken from the literature.  Our PDLA sample is
compared to the most complete sample of intervening DLAs currently
available.  Our principal conclusions are:

\begin{enumerate}
\item PDLAs exhibit a range of metallicities at a given redshift,
ranging from $\sim$ 1/3 to 1/1000 of the solar value (Section \ref{metal_sec}
and Figure \ref{Fe_Si_z}).  One of
the PDLAs in our sample exhibits the lowest N(SiII)/N(HI) of any known
DLA and has a value similiar to the intergalactic medium at this redshift.

\item Based on this modest-sized sample, there is a general trend
(with the exception of one fairly high velocity PDLA) for low
metallicities ($Z \sim 1/50 Z_{\odot}$) in PDLAs with log \nhi $<$ 21
and higher metallicities ($Z \sim 1/10 Z_{\odot}$) when the HI column
density is higher (Figure \ref{Fe_Si_S_Zn_HI}).  At these high \hi\
column densities, the metallicities of PDLAs are systematically higher
than the intervening sample by a factor of around three.

\item At least half of the PDLAs with \nhi\ $<$ 20.8 have sub-solar
ratios of [S/Si] which can not be easily explained by known dust or
nucleosynthetic trends (Section \ref{ion_alpha} and Figure
\ref{S_Si_HI}).  We interpret the low values as resulting from
ionization by a hard spectrum.  Sub-solar values of [S/Si] can be
present even in PDLAs with $\Delta V$ $>$ 2000 \kms, and with no
obvious trend with velocity.

\item In addition to the dependence of metallicity on \nhi\ and the
sub-solar [S/Si] ratios, other indications of enhanced/hard ionization
in the PDLAs which distinguish them from the intervening DLAs are: 1)
a possibly higher fraction of NV absorbers (tentative based on the
small number statistics of our sample), 2) higher fractions of
SiIV/SiII at low \nhi, 3) similar velocity structure in SiIV as SiII
in 3/5 QSOs where the comparison can be made, 4) low ratios of
[Ar/Si] (Section \ref{ion_sec}).

\item Most of the PDLAs (6/7) in our sample of seven have additional
high ionization gas at large negative velocities of $-100$ to $-825$
\kms\ (Section \ref{disc_high} and Table \ref{high_ion_tab}).  The
most extreme examples both have either a confirmed (Q0151+048) or
tentative (J1131+6044) close companion QSO.

\item One of the PDLAs in our sample has very sub-solar [Fe/Zn]
consistent with a large dust depletion fraction.  However, in general,
the [Fe/Zn] ratios of the PDLAs are consistent with the intervening
population (Section \ref{dust_sec} and Figure \ref{Zn_Fe}).  The PDLA
towards J1604+3951 has very different depletion in its two main
components.

\item The range of alpha-to-iron ratios in the PDLAs is also consistent with
the intervening DLAs, although there are a few notably high values (Section
\ref{alpha_sec} and Figure \ref{Si_Fe_HI}).

\item From analyses of the fine structure lines of SiII and CII,
QSO--absorber distances at least a few tens of kpc are determined,
with one case being constrained to have a separation of $>$ 160 kpc
(Section \ref{dist_sec}).

\end{enumerate}

We conclude that PDLA properties are generally consistent with an
origin external to the QSO host (in contrast with, e.g. the narrow
line associated systems, D'Odorico et al. 2004).  However, the larger
abundances (at least of high \nhi\ PDLAs) imply that they may not be
representative of the intervening sample.  We suggest that the PDLAs
may preferentially sample overdense environments where biased galaxy
formation has assembled more massive galaxies with higher
metallicities.  Indeed, at low redshift, some associated absorbers
have been identified with small clusters of galaxies (e.g. Bergeron \&
Boisse 1986; Hamann et al. 1997a).  If confirmed with larger larger
samples, this means that PDLAs could be used as probes of massive
galaxies at high redshift.  

PDLAs have been widely excluded from most DLA surveys.  The results
presented here indicate that, depending on the science objective,
this is a valid approach.  We have argued that PDLAs may sample
a rather special population of DLAs, possibly those in proto-clusters.
Although this renders the PDLAs an interesting probe of high
redshift galaxies, our results reveal some of the biases that could
be introduced into statistical surveys that do not impose a
$\Delta V$ limit in their DLA selection.  A high priority for
future work will be the improvement of emission redshift determinations.
The results presented here imply that even at fairly large relative
velocities ionization may affect abundance determinations.
Determining more accurate $\Delta V$ values in larger PDLA
samples will provide an empirically motivated cut-off for studies
of intervening DLAs.

\section*{Acknowledgments} 

We are grateful to the following people for sharing data that enabled
us to re-calculate emission redshifts for quasars in the literature:
Andrew Fox, Johan Fynbo, Isobel Hook, Michael Murphy, Samantha Rix and
Tayyaba Zafar.  Nikola Milutinovic and Nicolas Tejos helped with the
data acquisition of the HIRES and UVES data respectively.  Andrew Fox,
Palle M\o ller and Pasquier Noterdaeme provided useful comments on a
draft of the paper.  DMR acknowledges support from a Netherlands
Organization for Scientific Research (NWO) Veni Fellowship.

\end{document}